\documentclass[acmsmall]{acmart}

\AtBeginDocument{%
  \providecommand\BibTeX{{%
    \normalfont B\kern-0.5em{\scshape i\kern-0.25em b}\kern-0.8em\TeX}}}

\newcommand{\bheading}[1]{\vspace*{.5em}\noindent{\textbf{#1.}}}
\newcommand{\remove}[1]{}

\usepackage{pdfpages}
\usepackage{multirow}
\usepackage{soul}
\usepackage[normalem]{ulem}
\useunder{\uline}{\ul}{}
\usepackage[commandnameprefix=always, final]{changes}

\setcopyright{acmlicensed}
\acmJournal{PACMHCI}
\acmYear{2024} \acmVolume{8} \acmNumber{CSCW2}
\acmArticle{491} \acmMonth{11} \acmPrice{15.00}
\acmDOI{10.1145/3687030}

\begin{document}

\title[Conversational Agents for Deliberation of Harmful Content]{Conversational Agents to Facilitate Deliberation on Harmful Content in WhatsApp Groups}

\author{Dhruv Agarwal}
\email{da399@cornell.edu}
\orcid{0000-0002-1090-3583}
\affiliation{%
  \institution{Cornell University}
  % \streetaddress{P.O. Box 1212}
  \country{USA}
}

\author{Farhana Shahid}
\orcid{0000-0003-3004-7099}
\affiliation{%
  \institution{Cornell University}
  \country{USA}
}
\email{fs468@cornell.edu}

\author{Aditya Vashistha}
\orcid{0000-0001-5693-3326}
\affiliation{%
  \institution{Cornell University}
  \country{USA}
}
\email{adityav@cornell.edu}

% \affiliation{%

% \affiliation{%

% \affiliation{%

% \affiliation{%

% \affiliation{%

\renewcommand{\shortauthors}{Agarwal, et al.}

\begin{abstract}
WhatsApp groups have become a hotbed for the propagation of harmful content including misinformation, hate speech, polarizing content, and rumors, especially in Global South countries. Given the platform's end-to-end encryption, moderation responsibilities lie on group admins and members, who rarely contest such content. Another approach is fact-checking, which is unscalable, and can only contest factual content (e.g., misinformation) but not subjective content (e.g., hate speech). Drawing on recent literature, we explore deliberation---open and inclusive discussion---as an alternative. We investigate the role of a conversational agent in facilitating deliberation on harmful content in WhatsApp groups. We conducted semi-structured interviews with 21 Indian WhatsApp users, employing a design probe to showcase an example agent. Participants expressed the need for anonymity and recommended AI assistance to reduce the effort required in deliberation. They appreciated the agent's neutrality but pointed out the futility of deliberation in echo chamber groups. Our findings highlight design tensions for such an agent, including privacy versus group dynamics and freedom of speech in private spaces. We discuss the efficacy of deliberation using deliberative theory as a lens, compare deliberation with moderation and fact-checking, and provide design recommendations for future such systems. Ultimately, this work advances CSCW by offering insights into designing deliberative systems for combating harmful content in private group chats on social media.
\end{abstract}




%   \includegraphics[width=\textwidth]{images/Schematic Diagram.pdf}

\begin{teaserfigure}
    \includegraphics[width=\textwidth]{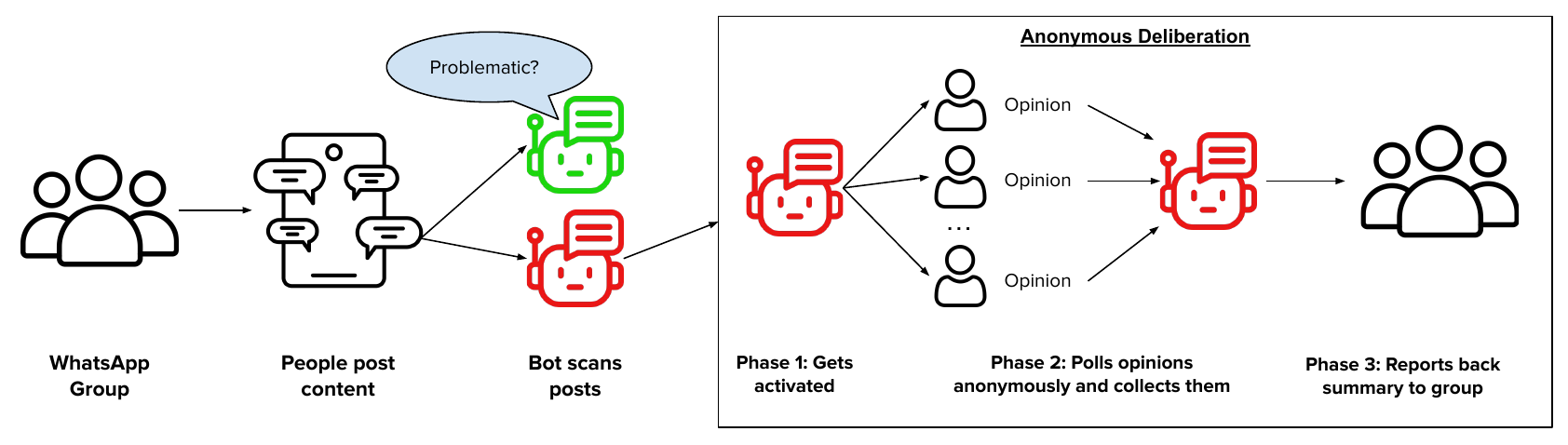}
    \caption{A schematic diagram showing the flow of the agent. It scans content in a group, gets activated on what it thinks is harmful, and then facilitates an anonymous deliberation process (shown in the box).}
    \label{fig:bot_schematic}
\end{teaserfigure}

\received{January 2024}
\received[revised]{April 2024}
\received[accepted]{May 2024}

\maketitle

\section{Introduction} \label{sec:introduction}

WhatsApp is a widely used instant messaging platform with billions of users worldwide. %Due to the high penetration of smartphones and WhatsApp's features, including end-to-end encryption, group chats, and flexible content modality %(text, images, audio, video) ~\cite{resende2019brazil, 2022strongties}, 
It has become the de facto standard for digital communication \remove{in many countries like India and Brazil }due to features such as end-to-end encryption, group chats, and multi-modality (text, images, audio, video)~\cite{resende2019brazil, 2022strongties}. 
%In recent years, 
However, the ease of communication in WhatsApp groups %have also emerged as a hotbed for 
has enabled harmful content such as misinformation, hate speech, rumors, and political or religious propaganda to proliferate quickly~\cite{resende2019brazil, Garimella2020}. %In particular, WhatsApp groups have been found to enable harmful content to go viral very quickly~\cite{resende2019brazil}. 
%The proliferation of harmful content on WhatsApp groups
This has contributed to election manipulation~\cite{FTWhatsappElection, Garimella2020, resende2019brazil}, deaths due to health myths~\cite{Islam2020}, and mob violence~\cite{Vasudeva2020lynchings} in many countries in the Global South.
% Combating such content is an important socio-technical problem of the information age~\cite{piccolo2021twitter, juneja2022human}.

A large body of HCI and CSCW scholarship has examined the spread of harmful content on WhatsApp~\cite{resende2019brazil, javed2021pakistan}, users' interactions with such content~\cite{2022accost, 2022usenix_students}, and approaches to mitigate its impact~\cite{kazemi2022tiplines, 2022strongties}. Two main mitigation strategies have emerged from this research: moderation and fact-checking. However, due to end-to-end encryption, WhatsApp cannot directly moderate content. Thus, the onus of identifying, managing, and removing harmful content falls primarily on admins and group members. Yet, admins rarely moderate content~\cite{FarhanaNewPaper}, and members refrain from contesting content to preserve group relationships~\cite{2022accost, 2022usenix_students}. WhatsApp has also established ``tiplines'' for users to request fact-checks from organizations in different regions~\cite{kazemi2022tiplines, MeedanTiplines}.
%fact-checking of user-reported content through ``tiplines'' has emerged as the only approach to identify harmful content on WhatsApp. %Tiplines are WhatsApp accounts that fact-check content for users; they have been found effective in capturing harmful content circulating on WhatsApp in a timely manner~\cite{kazemi2022tiplines}. 
But fact-checking is limited to debunking factual content such as misinformation and fake news, failing to address other types of harmful content that promote propaganda and polarization~\cite{juneja2022human}, which is popular in WhatsApp groups~\cite{resende2019brazil, javed2021pakistan, Garimella2020}. Further, users tend to keep fact-checks to themselves and do not share them in their groups to avoid confrontation with friends and family members~\cite{2022accost, 2022usenix_students, 2022strongties}. As a result, both moderation and fact-checking fail to combat harmful content in WhatsApp groups, making it an open and important challenge.

Recent scholarly work has introduced \textit{deliberation} as a useful approach to support accurate belief formation on social media~\cite{Bago2020deliberation}. Deliberation refers to the open and inclusive discussion of content seen on social media to make an informed decision about its legitimacy~\cite{jahanbakhsh2023personalized}. It nudges users to think critically about the content's accuracy and purpose~\cite{2021accuracy_nudges}. %This is in contrast to fact-checking where users are forced to believe a single source of truth~\cite{jahanbakhsh2023personalized}. %It is also privacy-preserving as it does not involve platform moderation. 
\citet{2022accost} demonstrated the effectiveness of deliberation in identifying COVID-19 misinformation in WhatsApp groups in rural India, though the deliberation occurred in offline social spaces rather than within WhatsApp groups. Parallelly, research has also shown the efficacy of conversational agents in facilitating discussions in online groups~\cite{2020botinthebunch, Kim2021moderator}. These studies have shown that such agents encourage deliberation and help surface diverse opinions.
%Deliberation was already seen in WhatsApp groups in India, where members would discuss COVID-19 misinformation amongst themselves~\cite{2022accost}. However, these deliberations would happen on side channels to avoid confrontation and would not make their way back to WhatsApp groups. 
Our work bridges these streams of work by examining the role of conversational agents in facilitating deliberation on harmful content in WhatsApp groups. %In this process, 
%In this work, we attempt to provide WhatsApp users a space to deliberate upon harmful content shared in their groups. Since conversational agents have been found to induce opinion deliberation in group chats~\cite{2020botinthebunch, Kim2021moderator}, 
%we also explore the design space of such an agent to facilitate the deliberation of harmful content in WhatsApp groups.
% We seek to understand what features could exist in the ultimate implementation of such an agent, even if it is not technically realizable today~\cite{indrani2017how}.
In particular, we ask the following research questions:
\begin{itemize}
    \item[\textbf{RQ1:}] How can we design a conversational agent to facilitate deliberation in WhatsApp groups?
    \item[\textbf{RQ2:}] Is community deliberation helpful in combating harmful content in WhatsApp groups?
\end{itemize}

To answer these questions, we used a design probe to conduct semi-structured interviews with 21 WhatsApp users in India. The probe showed an example conversational agent that would get activated upon detecting potentially harmful content in a WhatsApp group and then facilitate a deliberation process amongst the group members. We used this probe as a starting point for our conversation with participants.
% The agent functions in three steps: 1) It gets activated upon seeing problematic content in a WhatsApp group. 2) It then reaches out to group members via direct message (DM) asking them for their opinions about the content. 3) Finally, it collates responses from participants and reports them back to the group anonymously.
After participants engaged with the probe, we interviewed them to capture their perceptions around deliberation, including the strengths and limitations of using an agent to facilitate deliberation, their interaction with such an agent, and issues around privacy and group dynamics.
%benefits and limitations of an agent, how they would respond to it, and the trust and privacy implications of the agent. 
We ended the interview with a design worksheet which allowed the participant to customize the agent to their liking, thereby eliciting concrete feedback on its design features. 
% We found that participants received different kinds of problematic content like misinformation, hate speech, polarized speech, and financial fraud. They used various strategies to identify and respond to such content, including intuition and markers like emojis.
% They saw many advantages of the bot like its non-confrontational nature and intent to limit problematic content. They also identified limitations and suggested features like intelligence and privacy.

Overall, participants agreed that such an agent would help identify and contest harmful content in WhatsApp groups, but preferred anonymous deliberation to avoid confrontation with other group members. They preferred researching the content before expressing their opinions but were wary of the effort this required, recommending AI capabilities to reduce their workload. They appreciated the neutrality of the agent and found it useful for hearing diverse perspectives about a message. However, it could potentially disrupt group dynamics and would be futile for echo chamber groups. We also found several tensions in designing such an agent: calling out harmful content versus freedom of speech in a private space, giving an agent full access (privacy concern) versus having humans flag content (may disrupt group harmony), and the difficulty of including minority viewpoints. %opinion.
%Participants articulated several ways in which emerging AI capabilities can be integrated into the agent to facilitate deliberation with more ease. 
% We also found that participants favored researching before deliberating but expressed concerns about the effort this entailed. They suggested ways in which AI could reduce their workload and strategies to be selective about what content to discuss. They preferred anonymity in the deliberation process and were ambivalent about content moderation based on deliberation outcomes. Finally, we discuss the.  Some participants preferred simple content flagging and were unsure about its efficacy for extreme users.

Based on our findings, we discuss the efficacy of deliberation from the lens of deliberative theory and compare deliberation with existing approaches such as moderation and fact-checking. We conclude by offering concrete recommendations to design deliberative systems that help in identifying and combating harmful content in private group chats on WhatsApp. %For example, we recommend recruitment strategies, affordances for asynchronous discussion, the communication medium, human-AI collaboration, and anonymity in the deliberation process. We also propose using conversational agents as deliberative partners rather than mere facilitators.
Overall, we make the following contribution to CSCW literature:
\begin{enumerate}
    \item We explore the design space of using a conversational agent to facilitate the deliberation of harmful content in WhatsApp groups. We reveal design tensions in this space and show how users would interact with such an agent.
    \item We explore the efficacy of using deliberation to combat harmful content in WhatsApp groups, borrowing from relevant literature in deliberative theory. %In addition, we provide theoretical backing and the practical benefits of using this approach.
    \item We provide concrete recommendations to design deliberation systems for this purpose in the future.
\end{enumerate}

\section{Related Work} \label{sec:related_work}

%We aim to explore the design space of a conversational agent to facilitate the deliberation of harmful content on WhatsApp. 
%We ground our work in three categories of prior work. First, we discuss existing ways of combating harmful content on WhatsApp, establishing the utility of deliberation in this endeavor. We then strengthen our position by presenting supporting arguments from deliberative theory. Finally, we justify why a conversational agent is a useful interface for this purpose.

We first situate our work in emerging HCI and CSCW scholarship on identifying and addressing different types of harmful content on WhatsApp. We then present scholarly work on deliberative theory and its applications in facilitating deliberation and discussion in social media groups. Given our focus on deliberation through conversational agents, we then discuss studies in HCI and CSCW that use conversational agents in group communication. 

\subsection{Combating Harmful Content on WhatsApp}
A rich body of HCI and CSCW work has studied harmful content on WhatsApp, both qualitatively~\cite{2022accost, 2022usenix_students} and quantitatively~\cite{resende2019brazil, Machado2019Brazil, 2022strongties}. These studies have found the prevalence of various types of harmful content on WhatsApp, including political and religious propaganda and polarizing content~\cite{resende2019brazil, Machado2019Brazil, javed2021pakistan}, and misinformation, fake news, and rumors~\cite{2022accost, 2022usenix_students, javed2021pakistan}.

Scholars have also studied how WhatsApp users engage with harmful content and found that while many users can see such content as problematic, they hesitate to contest it openly in WhatsApp groups to avoid conflict with group members with whom they maintain social ties offline~\cite{2022usenix_students, 2022strongties}. Instead, group members often rely on credentialed group members, such as police officers and journalists, to discover, verify, and contest harmful content shared in WhatsApp groups~\cite{2022accost}. Given the lack of platform-mediated moderation on WhatsApp due to end-to-end encryption, WhatsApp users often rely on fact-checking through ``tiplines'' to access information credibility~\cite{kazemi2022tiplines, allen2021crowds}. Tiplines are platform-run or third-party WhatsApp accounts with which users share questionable content for fact-checking. %Thus, this approach provides on-platform fact-checking and bypasses end-to-end encryption since tiplines only fact-check content that users have self-reported.
However, despite their efficacy in debunking factual harmful content (e.g., misinformation and fake news) ~\cite{kazemi2022tiplines}, tiplines struggle with other kinds of harmful content that are subjective and opinionated (e.g., hate speech or polarizing content). Furthermore, fact-checking is often centralized, allowing either human fact-checkers~\cite{juneja2022human, haque2020factcheckers, nakov2021automatedfactchecking} or AI models~\cite{reis2019detection, ahmed2021detecting, mridha2021detection} to decide the truth~\cite{jahanbakhsh2023personalized}. This is at odds with the freedom of speech argument which demands that no centralized authority become the arbiter of truth~\cite{Jahanbakhsh2022Peer, koltay_protection_2022}. Indeed, users have expressed distrust in platform-mediated fact-checking due to platforms' political bias and for-profit policies~\cite{Saltz2021}. Moreover, centralized fact-checking faces scalability problems as there is far more content online than is possible for human fact-checkers to assess, especially when fact-checking cannot be fully automated~\cite{juneja2022human, allen2021crowds}. As a result, decentralized approaches to combating harmful content are gaining popularity in the research community~\cite{Jahanbakhsh2022Peer}.

Decentralized approaches aim to democratize moderation by allowing users to actively participate in the moderation process~\cite{jahanbakhsh2023personalized}. A prevalent approach is to appoint a set of users as moderators who are then responsible for identifying harmful content on the platform~\cite{Seering-2019, Cai-2022}, a strategy often used %to manage content 
on Reddit~\cite{Dosono2019RedditsMods} and Facebook groups~\cite{Kuo2023FacebookMods}. Another approach is crowdsourcing, which involves aggregating content ratings from a crowd of users~\cite{allen2021crowds, Pennycook2019CrowdsourceSources, Bhuiyan2020crowds, Bozarth2023RedditCrowds, Kim2018}. For example, ``Community Notes'' on X (Twitter) allows users to collaboratively add context to potentially misleading posts~\cite{twtter_community_notes}. While community moderation approaches are decentralized in the sense that users---not platforms---decide whether the content is harmful, they still impose a single source of truth (the moderators' or the crowd's decision) upon all users~\cite{jahanbakhsh2023personalized}.

Instead of imposing a singular authoritative source upon users, researchers have started proposing alternative strategies that empower users to independently assess content credibility~\cite{Jahanbakhsh2022Peer}, such as by incorporating credibility indicators~\cite{Lu2022CredibilityIndicators, Zhang2018CredbilityIndicators}, training personalized AI models to predict credibility~\cite{jahanbakhsh2023personalized}, and incorporating assessments from trusted peers~\cite{Jahanbakhsh2022Peer}. Deliberation is another decentralized approach that involves a discussion amongst people to gain a better understanding of an issue~\cite{Beauchamp2019} and has been shown to support accurate belief formation~\cite{Bago2020deliberation}. Like other decentralized approaches, deliberation allows users to discuss the content and empowers them to form their own beliefs about the quality of content. Unlike crowdsourcing approaches such as Community Notes on X (Twitter), the aim of deliberation is not to reach a consensus, but rather critical thinking and reflection~\cite{Bago2020deliberation}. Hence, no single opinion is forced upon users. Such decentralized approaches are more suitable for platforms like WhatsApp, where end-to-end encryption prevents centralized moderation. Yet, decentralized approaches have not been studied in the context of end-to-end encrypted platforms. We fill this gap by exploring the utility of deliberation to combat harmful content on WhatsApp.

\subsection{Deliberation and Deliberative Theory}
Deliberative theory, also known as deliberative democracy, is a political theory that emphasizes the importance of reasoned discussion, open dialogue, and inclusive deliberation in the decision-making processes of a society. Deliberative theory was developed by Habermas~\cite{Habermas1985} and Cohen~\cite{Cohen1989} in the 1980s in response to the limitations of traditional democratic theory~\cite{Beauchamp2019}. In particular, scholars argued that existing electoral systems, such as voting, may be fair and democratic, but do not produce decisions that best reflect the interests of participants~\cite{Bohman2000-BOHPDP, Ackerman2002-ACKDD, Mansbridge2010}. For example, voters may not know all the facts, may not have considered conflicting opinions, or may not have considered even their own opinions deeply~\cite{Fishkin2013}. On the other hand, deliberative theory argues that ``thoughtful, careful, or lengthy consideration by individuals''~\cite{Davies2012} promotes critical thinking and evidence-backed reasoning~\cite{Cohen1989}. Thus, deliberation allows people to ``better learn facts, ideas, and the underlying conceptual structures relating between them''~\cite{Beauchamp2019}, even under conditions of conflict and uncertainty~\cite{Friess2015}.

In recent years, deliberative theory has been applied to study and organize civic engagement online. For example, recent scholarly work on social media shows that the lack of critical thinking promotes belief in fake news, and deliberation results in the formation of more accurate beliefs~\cite{Bago2020deliberation, Pennycook2019lazy}. \citet{2022accost} reported collective deliberation practices within WhatsApp groups in India; urban users deliberated on potential misinformation in parallel ``satellite'' groups and rural users deliberated during in-person social gatherings. However, deliberation done through such informal structures often reinforces existing power structures and leads to poor deliberation outcomes. For example, the same study shows that marginalized members in rural communities could not express opinions during deliberation, and close ties and deference to elders constrained people from expressing their viewpoints in WhatsApp groups~\cite{2022usenix_students, 2022strongties, Scott2023Family}.

This is antithetical to the ideals of effective deliberation, which requires open discussion where individuals present and critically evaluate diverse viewpoints. It encourages inclusive processes that include diverse participants from various perspectives and backgrounds. Satisfying these ideals requires carefully considering three main components of deliberation defined by \citet{Friess2015}: the design, the process, and the outcome. Our work examines these three components of a conversational agent that creates a formal deliberative structure within WhatsApp groups to foster more informed, participatory, and reflective discussions on harmful content. We now discuss prior HCI and CSCW on using conversational agents to facilitate discussions in online groups.

%These discussions were done through informal structures without any platform support which led to poor deliberation practices. For example, in rural areas, marginalized identities were unable to express opinions~\cite{2022accost}, and in urban areas, close ties amongst family members prevented participants from expressing their opinions~\cite{2022usenix_students, 2022strongties, Scott2023Family}. We extend this prior work by systematically exploring the design space of enabling deliberation in WhatsApp groups.

% ### moved this para to the beginning of findings
% To explore the design space of conversational agents for this purpose, we utilize \citet{Friess2015}'s framework to design online deliberation systems. According to this framework, there are three main components of deliberation: the design, the process, and the outcome. The design includes the institutional framework under which the deliberation takes place, such as the environment that enables and fosters deliberation~\cite{Beauchamp2019}. The process encompasses the actions of participants and the quality of communication. And the outcome focuses on the (expected) results of the deliberation. We organize our findings using this framework to comprehensively evaluate the design space.  We now discuss prior HCI and CSCW on using conversational agents in group settings.  

\subsection{Conversational Agents in Groups}
Conversational agents or chatbots %\footnote{We label our probe as a ``conversational agent'', avoiding the term ``chatbot'' due to its recent association with generative capabilities, as seen in examples like ChatGPT.} 
are computer programs designed to converse with humans using natural language~\cite{Hussain2019survey}. They have been increasingly used to help maintain, moderate, or grow online communities~\cite{Seering2018socialbots}. A review by \citet{Seering2019dyadic} found that most chatbots are dyadic: they support one-on-one user interactions and are either task-oriented or chat-oriented. Task-oriented chatbots perform a task or provide a service to the user upon their request, such as providing customer service~\cite{Xu2017customer} or tracking nutrition~\cite{Graf2015nutrition}. Chat-oriented chatbots are more performative and are intended to showcase advanced natural language processing and generation capabilities. However, such dyadic chatbots rarely support group interactions.
% These dyadic chatbots rarely support group interactions.
% Recently, general-purpose chatbots like OpenAI's ChatGPT, Google's Bard, etc. have blurred these boundaries by providing services while simultaneously showcasing natural language abilities.
% Past work has focused on the design and perceptions of such dyadic chatbots that rarely support group interactions~\cite{Seering2019dyadic}.

In response, recent work has started to propose multi-party chatbots that communicate with multiple users in a group. For example, researchers built bots to mediate task management in teams~\cite{Toxtli2018ChatbotTaskManagement}, engage in a coffee chat with humans~\cite{Candello2018ChatbotCoffee}, and facilitate online discussions~\cite{Lee2020solution, 2020botinthebunch}. Facilitating discussions has emerged as an especially important application of multi-party chatbots to encourage rational debate, which is often lacking in group chats on platforms such as WhatsApp and Telegram~\cite{Kim2021moderator}. This is due to a variety of reasons. First, group chats, unlike structured forums, are fast-moving, unstructured, and unthreaded, which makes it difficult for users to keep track of diverse opinions expressed in the group~\cite{2020botinthebunch}. Second, group chats are associated with procrastination, loss of concentration, and uneven participation~\cite{Duin1996CollaborationVE, 2020botinthebunch, Schultz2000OnlineForums}. Thirdly, people may not be accepting of contrasting opinions in online groups~\cite{Mutz2006hearing}. To overcome these problems, recent work has proposed using chatbots to facilitate effective deliberation in group chats~\cite{2020botinthebunch, Kim2021moderator}. Integrating chatbots into users' familiar environment rather than building external interfaces has proven useful in encouraging deliberation, facilitating decision-making and open-debating, and helping surface diverse opinions~\cite{Kim2021moderator, 2020botinthebunch}.

However, these works have only considered discussions about non-sensitive topics such as travel decisions, ethical dilemmas, or estimation (e.g., height of the Eiffel Tower). In fact, \citet{Kim2021moderator} recognized the need for ``specialized interventions'' for deliberation about sensitive or divisive topics that could be affected by social dynamics. We extend this line of work by exploring the design of an agent to facilitate the deliberation of harmful content in WhatsApp groups, a potentially divisive topic affected by social, cultural, and power dynamics~\cite{2022accost, 2022usenix_students, 2022strongties}. 

\section{Methodology} \label{sec:methodology}

To answer our research questions, we conducted a qualitative study with 21 WhatsApp users in urban and rural India. Our participants were part of multiple WhatsApp groups and encountered harmful content such as hate speech and misinformation in their groups. Participants first engaged with a design probe showcasing a conversational agent for facilitating deliberation in WhatsApp groups, followed by their participation in a semi-structured interview.
% Our participants first engaged with a design probe of a conversational agent that can facilitate deliberation in WhatsApp groups and then participated in a semi-structured interview.

%In this section, we describe the different components used in our study.
% We conducted the interviews %were conducted both 
%either via Zoom or in-person to accommodate the geographical spread of our interviewees. %The study was conducted in May--June 2023 and the data was analyzed thereafter.

\subsection{Design Probe}

\begin{figure}[t]
    \centering
    \begin{tabular}{ccc}
    \includegraphics[height=2.2in]{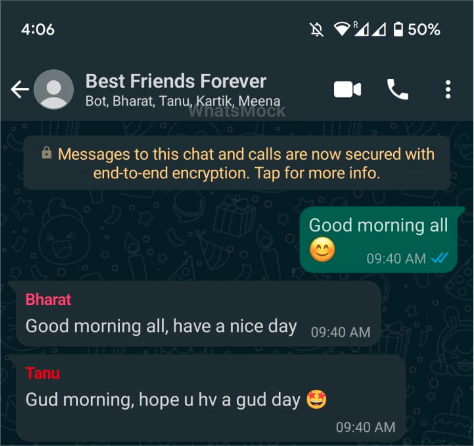} & \includegraphics[height=2.2in]{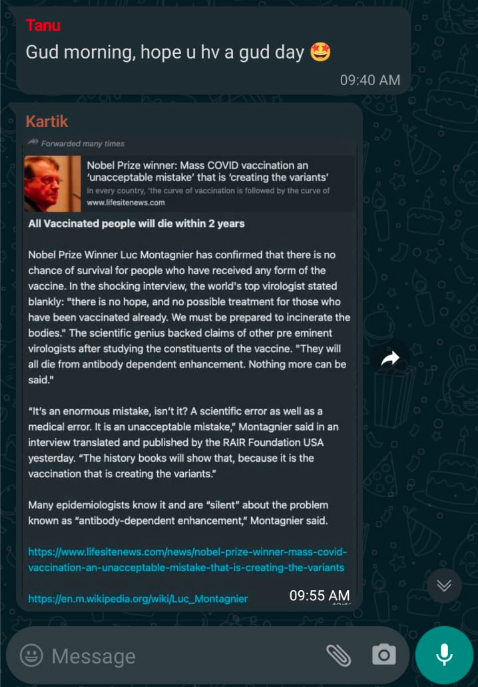} \\
    (a) Setting the context & (b) Misinformation shared \\
    \end{tabular}
    \caption{The probe showed a fictional informal WhatsApp group among friends where sometimes harmful content is shared. The agent is shown in Figure~\ref{fig:bot_design} in the appendix.}
    \label{fig:probe_design}
\end{figure}

HCI researchers often use design provocations to understand the needs and desires of users in real-world settings and get feedback on new technologies~\cite{hutchinson_technology_2003}. 
%depict abstract concepts to users. 
%In our study, we adopted design probes to examine the design of a conversational agent to facilitate the deliberation of harmful content in WhatsApp groups. 
Since our participants might not be familiar with the idea of using conversational agents to facilitate online deliberation, we created a design probe to show participants an example of how such an agent could be implemented. % with which participants could interact. The tangible interaction with the probe
Seeing a tangible example allowed participants to better comprehend and engage with the concept. This allowed them to think critically about the capabilities, strengths, and limitations of using a conversational agent to deliberate on harmful content in WhatsApp groups. The probe served as a starting point for our discussion with the participants.

We designed the probe using a free WhatsApp mock chat creator called WhatsMock~\cite{whatsmock}.
%During the interview, we first set the context by showing participants a hypothetical group chat titled ``Best Friends Forever'' (Figure~\ref{fig:probe_design}). We emphasized the informal nature of the group by showing fictional group members sharing greetings and pleasantries. Next, we showed one of the group members forwarding vaccine-related misinformation which said, ``\textit{All vaccinated people will die within 2 years.}'' We sourced this message from a popular fact-checking website in India that had gone viral on WhatsApp in May 2021. The message was tagged as `forwarded many times', a label introduced by WhatsApp in 2019 to alert users and slow down the spread of harmful content~\cite{forwarding-limits}. Following this, we introduced different aspects of the agent, which are explained below.
The probe first showed a fictional WhatsApp group titled ``Best Friends Forever'' (see Figure~\ref{fig:probe_design}) in which group members shared greetings, pleasantries, and information with each other. The probe then showed one of the group members forwarding a message that was tagged as `forwarded many times'\footnote{`Forwarded many times' is a label introduced by WhatsApp in 2019 to alert users and slow down the spread of harmful content~\cite{forwarding-limits}} and said, ``\textit{All vaccinated people will die within 2 years.}'' We curated this message from a popular fact-checking website in India that debunked it as misinformation. %, went viral on WhatsApp in May 2021. 
% The message was tagged as ``forwarded many times'', a label introduced by WhatsApp in 2019 to alert users and slow down the spread of harmful content~\cite{forwarding-limits}. 
The probe then showed an example design of a conversational agent integrated into the WhatsApp group to facilitate the deliberation of potentially harmful content, so that participants think about the design critically and concretely instead of abstractly. 

% ### moved to the first page of the paper
% \begin{figure}[t]
%     \centering
%     \begin{tabular}{c}
%     \includegraphics[width=\textwidth]{images/Schematic Diagram.pdf}
%     \end{tabular}
%     \caption{A schematic diagram showing the flow of the agent. It scans content in a group, gets activated on what it thinks is harmful, and then facilitates an anonymous deliberation process (shown in the box).}
%     \label{fig:bot_schematic}
% \end{figure}

\begin{figure}[t]
    \centering
    \begin{tabular}{ccc}
    \includegraphics[height=2.4in]{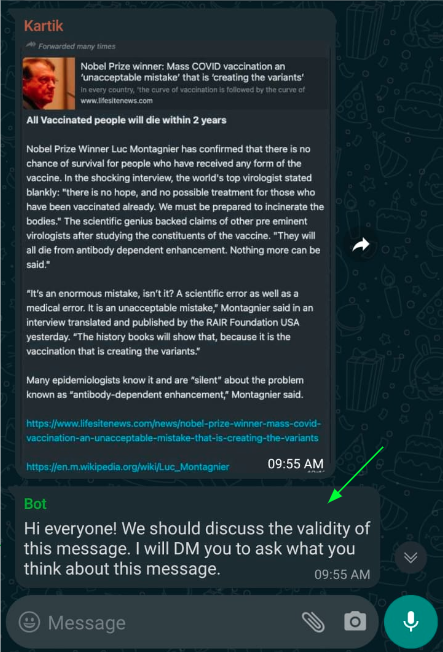} & \includegraphics[height=2.4in]{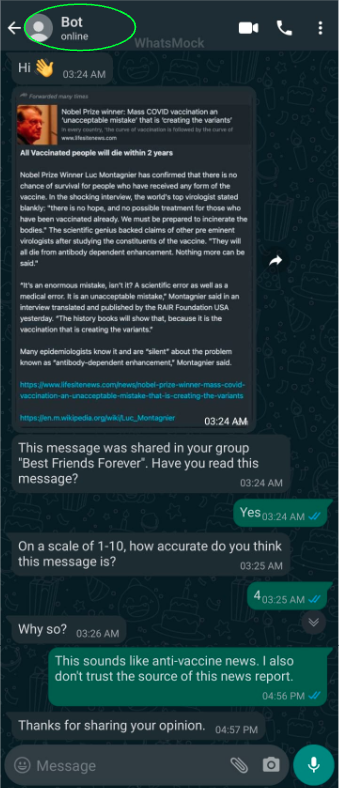} & \includegraphics[height=2.4in]{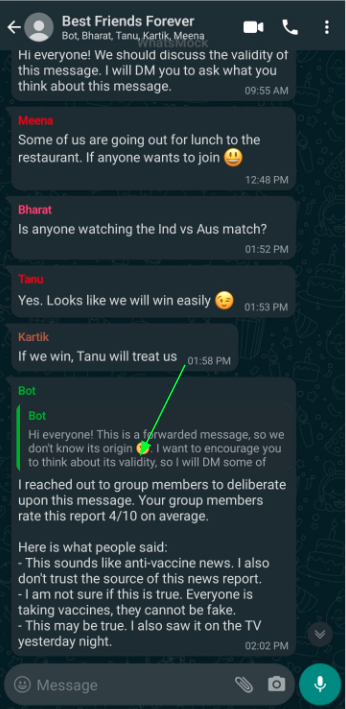} \\
    (a) Activation & (b) Polling opinions & (c) Summarization  \\
    \end{tabular}
    \caption{Different phases where the agent interacts with WhatsApp group members.}
    \label{fig:bot_design}
\end{figure}

\subsubsection{Design of the Conversational Agent} \label{subsec:bot_design}
Our example probe showed the conversational agent as a member of the group with the ability to see messages in the group, send direct messages (DMs) to other users, and have a contact name and a profile picture. We did not give the agent a name or profile picture to avoid biasing participants about its functionality. Borrowing from the literature on facilitation bots~\cite{2020botinthebunch}, the agent divided the deliberation process into three phases:
%initiating the deliberation, the deliberation itself, and summarizing what was discussed. 
%initiating the deliberation, polling opinions for deliberation, and producing a summary for deliberation. %In terms of the agent, these steps translated into three phases:
\begin{enumerate}
    \item \textbf{Activation:} The agent gets activated upon seeing harmful content and initiates the deliberation process.
    \item \textbf{Polling Opinions:} The agent encourages group members to reflect on the harmful message and collects their opinions via direct message (DM).
    \item \textbf{Summarization:} The agent collates the opinions into an anonymous summary and shares it back in the group for members to reflect upon and deliberate on collectively.
\end{enumerate}
Figure~\ref{fig:bot_schematic} shows the schematic representation of the three phases and Figure~\ref{fig:bot_design} shows an example scenario of the agent functioning in a group.

Overall, our agent design encouraged deliberation in two ways. First, it solicited anonymous opinions from group members. This facilitated idea-sharing even in the presence of close ties within the group~\cite{2022usenix_students, 2022strongties}, thereby prioritizing inclusivity---an essential aspect of deliberation theory~\cite{Fishkin2013, Habermas1985}. Second, it posted a summary message containing everyone's opinion. This prompted reflection amongst group members, setting the stage for further deliberation and fostering open discussion within the group. We now explain the three phases listed above in detail.

\bheading{Activation}
The agent gets activated upon seeing harmful content in the group and sends the following message in the group: \textit{``Hi everyone! We should discuss the validity of this message. I will DM you to ask what you think about this message.''}
As prior research shows that nudging people to think critically about information significantly reduces its propagation~\cite{2021accuracy_nudges}, we showed that the agent sends this message as an accuracy nudge instead of silently polling users in the background. However, since this was only a design provocation, we purposely left out many details so that participants could provide suggestions about different aspects of the activation process.

\bheading{Polling Opinions} The agent then reaches out to the group members via direct message (DM). It greets the user and sends them the harmful message upon which the agent wants reflection and deliberation. After confirming if the user has seen the message, it prompts them to (a) rate the accuracy of the message on a scale of 1--10, and (b) provide reasoning behind their rating. The rating was intended to encourage participants to assess the message critically. Even if the content was factually correct (e.g., some hate speech may not be fabricated), assigning a rating could aid in recognizing its harmful nature.
% We chose to use a scale instead of a binary true-false rating based on previous work that shows fact-checking is not black and white and involves a lot of gray areas~\cite{juneja2022human}.
The agent also asks for the reasoning behind their rating to further encourage reflection as prior research shows that asking users to reason about their rating encourages them to think more deeply about the content~\cite{jahanbakhsh2023personalized}. After the user sends their response, the agent thanks them for sharing their opinion.
%. The framing of this message subtly reinforces that to indicate that this is an \emph{opinion}-based deliberation process.

\bheading{Summarization} The agent collates the responses received from multiple group members and reports them back to the group in the form of a summary message. In the summary message, the agent reports the average accuracy rating and lists the collected opinions anonymously. We listed opinions anonymously because prior work highlights the importance of anonymity in sharing opinions on sensitive content~\cite{2022usenix_students, 2022accost}, and this also enabled us to examine participants' preferences around anonymity in the deliberation process. The agent then encourages the group members to collectively deliberate on the shared summary.
% to avoid awkwardness in correcting close ties publicly, as reported by past work~\cite{2020botinthebunch}.
%Also, to highlight that the deliberation process takes time, the probe showed the summary message being sent by the agent four hours after the harmful message was sent. % and the informal banter in the group before the summary message.
To emphasize the time-consuming nature of the deliberation process, the probe showed that the summary was sent four hours after the original harmful message (see Figure~\ref{fig:bot_design}c). While this may risk resurfacing a harmful message in the group, it has been found that harmful content left unaddressed can falsely imply to users that it's not harmful~\cite{Pennycook2020Implied}. Thus, it is important to call out harmful content it even after a delay.

% \subsection{Probe Design} \label{subsec:probe_design}
%Since a bot is a very dynamic entity, initially we found it hard to demo a bot prototype to participants. We considered creating a product video explaining the bot's features, but through discussions within the research team, we realized that people use WhatsApp in different ways, so a hard-coded video would be difficult to understand for diverse users. Hence, we decided to demo the prototype by using screenshots of the bot in a PowerPoint presentation with short captions alongside each screenshot. This would allow the interviewer to explain each step of the bot based on each participant's context, and conversely, it would allow participants to ask questions during the demo rather than having to wait until the end of a video.

\subsection{Semi-structured Interviews}

\begin{figure}[t]
    \centering
    \begin{tabular}{ccc}
    \includegraphics[height=1.45in]{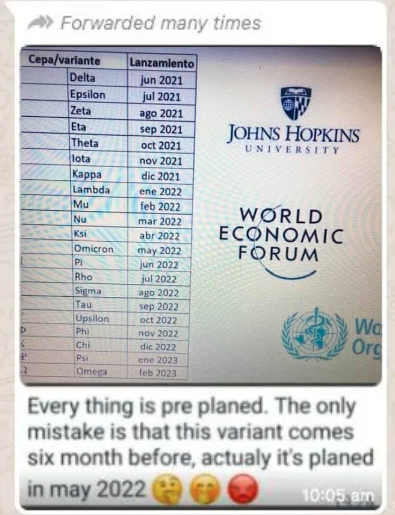} & \includegraphics[height=1.45in]{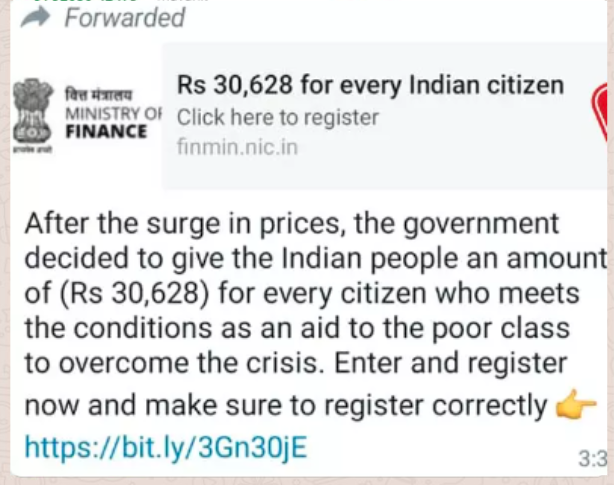} & \includegraphics[height=1.45in]{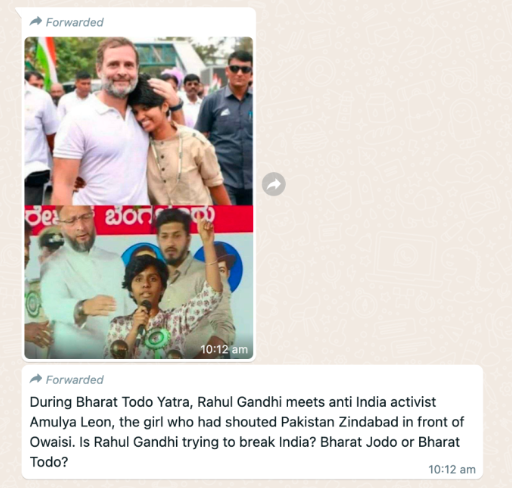} \\
    (a) COVID-19 misinformation & (b) Financial misinformation & (c) Polarizing content \\
    \end{tabular}
    \caption{After demoing the agent, participants were shown one of these three harmful messages. They were asked the same questions the agent would ask them: to rate the accuracy of the message on a scale of 1--10 and provide reasoning behind their rating. The three messages were sourced from popular fact-checking websites in India.}
    \label{fig:problematic_messages_shown}
\end{figure}

Once participants engaged with the design probe, we conducted semi-structured interviews with them to understand their needs, preferences, and concerns about using an agent to deliberate on harmful content. We conducted 21 interviews, stopping when the responses reached theoretical saturation. Each interview lasted for about 60--90 minutes and was conducted either in-person or virtually over Zoom in Hindi or English based on participants' preferences. %We paid participants INR 300 ($\approx$ \$3.67) in the form of gift coupons for participating in the interview. Each interview broadly consisted of three segments: demographic questions for context, a demo of the probe and questions about it, and finally a design worksheet.
Each interview started with a brief introduction to the study and the informed consent process. After receiving consent, we first asked questions to understand the demographics of the participants, their use of WhatsApp, and their experiences with harmful content in WhatsApp groups.
%In the next segment, we introduced the design probe by taking them through the different phases of the agent. %the presentation. 
%We encouraged them to critically engage with it by emphasizing that this was merely a design suggestion and not the final design.

Subsequently, we asked them questions to understand their preferences for different design features, the perceived benefits and limitations of the deliberation process, and concerns around embedding an agent in WhatsApp groups. We then showed them a harmful WhatsApp message and asked how they would respond if the agent asked for their opinion on this message. Instead of showing the same message to all participants, we showed them one of the three messages sourced from fact-checking websites in India: a COVID misinformation, phishing scam, and politically polarizing content (see Figure~\ref{fig:problematic_messages_shown}). 
%. One of these was COVID misinformation easily distinguishable with spelling errors, the use of emojis, and a blurry image~\cite{2022usenix_students}; the other was financial misinformation with an unverifiable link but slightly harder to distinguish due to a government logo; the last was polarizing political misinformation that was very hard to distinguish with no obvious credibility indicators. All these messages had the ``forwarded'' tag, a feature on WhatsApp that alerts users about forwarded content.
We then asked participants questions about trust, privacy, and the impact of deliberation on group dynamics.

The final segment of the interview was a design worksheet titled ``Let's Design this Bot Together!''\footnote{The worksheet is attached in Appendix~\ref{sec:design_worksheet}.} which the participants filled. In virtual interviews, the participant shared verbal responses and the interviewer filled out the worksheet while sharing their screen.
% We handed the participants a pen to fill out the worksheet and discussed their responses as they did so.
The worksheet contained questions about design decisions associated with the agent, such as possible activation strategies, how many people the bot should reach out to, and how long it should wait for people's opinions, among others. While some of these questions had already been discussed by this point in the interview, the worksheet encouraged participants to carefully consider tradeoffs in different design choices and to clearly communicate their design preferences. 
%to concretize design preferences. 
%design decisions and consider the design more deeply.
% Since past work has shown the importance of the personality of a bot~\cite{indrani2017how, 2018farmchat}, the worksheet also had questions about the personality such as giving it a name, profile picture, gender, etc. However, we did not receive rich responses to these questions in the initial interviews, so we removed these questions from the remaining interviews. This may have happened because users expect a bot's personality to match its domain~\cite{jain2018evaluating}, so they wanted a misinformation-related bot to maintain a neutral personality.
Finally, we thanked the participants for their time and gave them a gift card of
INR 300 ($\approx$ \$3.67). %in the form of gift coupons for participating in the interview. 
%answered their questions, if any, to end the interview.

\subsection{Analysis}
All interviews were recorded with the consent of the participants. We transcribed the interviews, translating them into English if required. This resulted in around 20 hours of audio recordings and 190 pages of transcription. The first author then analyzed the transcripts using inductive thematic analysis. During this process, all co-authors met periodically to refine the codes, merge similar ones, stabilize the codebook, and discuss emerging themes. After multiple iterations through the data, we ended up with $156$ codes, which we categorized into eight sub-themes such as activation of the agent, perception of the agent, and deliberation as a strategy. Finally, we mapped these sub-themes into the three components of deliberation defined by \citet{Friess2015}: design, process, and outcome. Thus, we used a hybrid inductive-deductive approach, formally referred to as abductive analysis~\cite{VilaHenninger2022, Timmermans2012}.

\subsection{Participants}
We used a combination of convenience and snowball sampling to recruit participants. We started by advertising the study through our immediate network and interviewed people who expressed interest in participating in our study. We also partnered with a non-profit organization based in rural India to recruit WhatsApp users in rural areas. %We then asked our participants to connect to their network. Throughout this process, we made a conscious effort
We tried to recruit participants of different genders, age groups, and urbanity considering that people with different demographic backgrounds have different experiences with harmful content~\cite{2022accost, 2022usenix_students}.

\begin{table}
\centering

\begin{tabular}{|l l l l l | l l l l l |}
\hline
\textbf{ID} & \textbf{Gen}& \textbf{Age} & \textbf{Edu}& \textbf{Occupation} & \textbf{ID} & \textbf{Gen}& \textbf{Age} & \textbf{Edu}& \textbf{Occupation} \\ \hline
P1 & F & 27 & M & Technician & P12 & M & 28 & B & Student \\
P2 & M & 24 & B & PhD Student & P13 & F & 40 & M & NGO worker \\
P3 & M & 26 & M & Auditing analyst & P14 & F & 25 & M & Coaching teacher \\
P4 & F & 23 & B & Research Fellow & P15 & F & 39 & HS & Housewife \\
P5 & F & 21 & HS & Student & P16 & M & 26 & B & Factory worker \\
P6 & M & 24 & B & Business Consultant & P17 & M & 20 & MS & Electrician \\
P7 & M & 30 & B & Product Manager & P18 & F & 42 & M & Housewife \\
P8 & F & 23 & B & Student & P19 & F & 29 & M & Village leader \\
P9 & F & 28 & M & Consultant & P20 & M & 32 & B & Teacher \\
P10 & F & 27 & B & Product Designer & P21 & M & 32 & B & Factory worker \\
P11 & F & 24 & B & Creative Associate &  &  &  &  &  \\\hline
\end{tabular}

\caption{Participant ID, gender (M: male, F: female), age, educational level (M: master, B: bachelor, HS: high school, MS: middle school), and occupation of all 21 participants. All details are self-identified.}
\label{tab:participant-demographics}

\end{table}

%We recruited 21 Indian WhatsApp users for our study, who were part of different WhatsApp groups and encountered harmful content such as hate speech and misinformation in their groups.

\bheading{Demographics}
Table~\ref{tab:participant-demographics} shows our participants' demographic details. In total, our study had 21 participants, of which $12$ identified as female and the rest as male. The participants were in the age range of $20$--$42$ years with an average age of $28.1$ years.  All of them owned a personal smartphone on average for $7.3$ years and used WhatsApp on average for $6.9$ years. Our participants had varying literacy levels: $3$ completed high school or lower, $11$ had a bachelor's degree, and $7$ had a master's degree. Some of them were students and homemakers and others worked as technicians, finance analysts, NGO workers, and managers. There was an even split of urbanity: $11$ lived in urban areas and $10$ lived in peri-urban or rural areas.
% While some rural participants had moved to urban areas over time, their families and friends with whom they were in WhatsApp groups still lived in rural areas.

\bheading{Experience with WhatsApp}
Participants extensively used WhatsApp to maintain different aspects of their daily lives, such as work, education, and personal connections, primarily through participation in different groups. These included social (friends, family, village, college), professional (e.g., job opportunities), educational (e.g., college classmates), and organizational (religious, political, workplace) groups. %Rural users also engaged in media groups to share local news. 
Group sizes ranged from 4 to 200 members. Social groups of friends and family were smaller in size (4-40 members) and those comprising village residents or college classmates were larger (up to 200 members). The size of the organizational and professional groups depended on the scale of the organization, with political or religious groups being mid-sized (20-50 members). In small groups, members typically knew each other, while larger groups often comprised geographically dispersed members with weak social ties.

\subsection{Ethics and Positionality}
This study was approved by the IRB of our institution. We took several steps to conduct research ethically and responsibly. For example, recognizing the risks associated with deploying an agent in private online spaces, we opted to use a design probe instead of other methodologies (e.g., Wizard-of-Oz, working prototype) to protect the privacy of our participants. Additionally, to avoid the risks of propagating harmful content through this study, we clarified that the examples shown in the study are misinformation and should not be passed on to peers.

%Our team consists of researchers from countries in the Global South and has rich experience conducting qualitative work with rural and urban communities in India. Hence, we are familiar with the local culture and social norms of the region. We have also personally experienced harmful content in various WhatsApp groups we are part of. This cultural proximity was valuable in practice. It allowed us to identify diversity in participants' opinions, create culturally relevant design artifacts, and form a rapport with the participants before diving into the interview. However, we acknowledge that our academic training and urbanity would have affected the analysis and synthesis of this work.

Our research team consists of authors who have lived in India for a considerable amount of time. We also have a decade-long experience in conducting research with different rural and urban communities in India and other countries in the Global South. As a result, our team had a good understanding of local sociocultural norms, sociopolitical landscape, and sociolegal frameworks. This helped us not only build rapport with the participants but also understand the context within which harmful content propagates in online communities in India. This cultural proximity also enabled us to create culturally relevant design artifacts.
% and understand the factors that impacted participants' engagement with harmful content in WhatsApp groups.
We approached this work from an emancipatory action research mindset to understand users' experiences with deliberation in end-to-end encrypted online communities like WhatsApp groups. We aimed to inform design recommendations to build conversational agents that facilitate deliberation on divisive topics in an understudied context where sociocultural norms and power dynamics influence how people engage with and contest harmful content online.

% \section{Landscape of WhatsApp Groups} \label{sec:landscape}
% \input{4 landscape}

\section{Findings} \label{sec:findings}

%To explore the design space of conversational agents for this purpose, we utilize \citet{Friess2015}'s framework to design online deliberation systems. According to this framework, there are three main components of deliberation: the design, the process, and the outcome. The design includes the institutional framework under which the deliberation takes place, such as the environment that enables and fosters deliberation~\cite{Beauchamp2019}. The process encompasses the actions of participants and the quality of communication. And the outcome focuses on the (expected) results of the deliberation. We organize our findings using this framework to comprehensively evaluate the design space.  

%We organize our findings according to the three dimensions of deliberation defined by scholars of deliberative theory~\cite{Friess2015, Beauchamp2019}. 
As discussed in Section~\ref{sec:related_work}, we organise our findings using the three main components of deliberation presented by \citet{Friess2015}: the design, the process, and the outcome. The \textit{design} (Section~\ref{subsec:environment}) includes the institutional framework under which the deliberation takes place, such as the environment that enables and fosters deliberation~\cite{Beauchamp2019}. The \textit{process} (Section~\ref{subsec:anon_deliberation}) encompasses the actions of participants and the quality of communication. And the \textit{outcome} (Section~\ref{subsec:is_anon_delib_helpful}) focuses on the (expected) results of the deliberation. We use this framework to comprehensively evaluate the role that conversational agents can play in the deliberation of harmful content in WhatsApp groups.

\subsection{Design: The Deliberative Environment} \label{subsec:environment}
%Below, we focus on three main aspects of our agent's design: activation, participation, and duration. We asked participants about all these aspects; we report the findings below. The key findings from this section are summarized in Table~\ref{tab:design-key-findings}.
Our participants shared several preferences about the design of the agent, including how it should get activated and initiate deliberation in WhatsApp groups, who it should invite to participate in the deliberation, and how long people should deliberate. 
%what should be the mechanics of deliberation.

% \input{tables/design_findings_summary}

% We first report participants' perceptions and preferences for how and when the agent should be activated.
% Below, we describe the activation strategies that surfaced in our conversations with participants.
% Once activated, the agent would decide who would participate in the deliberation. Hence, we discuss who the agent should reach out to and how long the deliberation should last.

\subsubsection{Initiating the Deliberation} \label{subsubsec:activation}
% To use an agent for any purpose, a fundamental decision must be made about when and how the agent will get activated.
While describing the agent, we told participants by way of example that the agent would get activated whenever any harmful content would be shared in the group. However, we did not specify details about the activation process, for example, how the agent would know whether a message is harmful or not. This ambiguity encouraged participants to reflect on different ways in which an agent can be activated to initiate deliberation and the resulting tradeoffs. In our analysis, three broad activation strategies emerged: heuristics-based, AI-based, and manual activation.

\bheading{Heuristics-Based Activation}
Participants initially proposed that the agent should monitor content and get activated automatically when it detects certain indicators of harmful content. For example, they suggested the agent could use WhatsApp-provided tags (``forwarded'' or ``forwarded many times'') to detect harmful messages since these tags were considered a ``\textit{hallmark of harmful content}'' (P2). Others suggested using features such as message length, embedded links, emojis, and font styles (e.g., excessive use of bold or italics) to detect potentially harmful content. Some participants suggested that the agent should get activated on ``suspicious'' phrases such as slurs (e.g., misogynistic or racial terms), political keywords (e.g., war, elections), or words commonly appearing in monetary scams (e.g., ``send money'').

However, some participants pointed out that these heuristic-based methods were susceptible to false positives and may end up flagging content that may not be harmful. % While that may be true in general, an example Participants' comments raised important design tension between freedom of speech and calling out harmful content. 
For example, P5 worried that if the agent detects keywords, it might flag group members' personal opinions instead of ``external'' content (e.g., news reports), suggesting that users may not perceive personal opinions as harmful. P5 elaborated: 
\begin{quote}
    ``\textit{If I say a racist or homophobic thing, it can be my opinion...it may not be news. So it will get activated on that...and people will feel that their free speech is curtailed.}'' % -- P5
\end{quote}
Thus, users preferred using the agent to deliberate on forwarded messages instead of their ``personal opinions'' even though they may be toxic or hateful towards others. This raises an important design tension between maintaining freedom of speech in a private space and calling out harmful content in the group. \label{design_tension:freedom_of_speech}
% This suggests that users may not classify group members' personal opinions as harmful. 
% This raises the question: Should harmful content (e.g., hate speech) be called out when coming directly from a group member? Thus, there is a design tension between freedom of speech and calling out harmful content. 

\bheading{AI-Based Activation}
Some participants wanted the agent to be intelligent enough to detect various forms of harmful content, including graphic images, sensitive content, non-credible sources, and harmful audio/video messages. P16 stated: %In general, they expected the agent to able to identify harmful content.
\begin{quote}
    ``\textit{There must be something in the app through which it will know [what is harmful], the agent must have some capability.}'' %-- P16
\end{quote}

Upon receiving these suggestions, we asked participants what was the need for deliberation if we already possessed an omniscient agent that could identify all harmful content. After reflection, they recognized the challenges in building such an agent and adjusted their expectations accordingly. P8 suggested monitoring the tone of conversations in the group and activating the agent when the group fell into a fight or argument. Others suggested using historical evidence to identify harmful content. For instance, P1 proposed that the agent could use group members' opinions to learn the features of harmful content over time, an approach similar to that used by \citet{jahanbakhsh2023personalized}. P1 elaborated: 
\begin{quote}
    ``\textit{Will the agent be learning all this while?... Will it be able to filter out misinformed content in the future instead of reaching out every time? ... That means it won't message people asking for their opinion every time.}'' %-- P1
\end{quote}
P1 argued that this would allow the agent to identify harmful content autonomously, ultimately reducing the frequency of the deliberation process over time. %This approach is interesting because it trains a new classifier (to identify harmful content) within each group instead of a global classifier that would undermine the diversity of opinions on WhatsApp.
% provide direct verdicts instead of beginning a deliberation process.

This idea exemplified a general suggestion of human-AI collaboration where the agent would reach out to participants only when it couldn't identify harmful content autonomously, thereby reducing their workload. Another such example came from P19 who suggested that the agent could find relevant sources (e.g., news articles, fact-checks, etc.) and compare the message against them to gauge veracity before reaching out for opinions.
\begin{quote}
    ``\textit{This agent will work better if it could search for an authentic source on its own, be it any report from the government or international organization or a standard newspaper... So if it could match the news with any such sources and back its existence. That is better than taking people's individual opinions. That would be more authentic.}'' -- P19
\end{quote}

Both the above activation strategies---heuristics-based and AI-based---were automatic: the agent would have to ``see'' or ``scan'' all messages in the group to get activated. This was perceived as a privacy concern in smaller groups, where participants didn't want the agent to monitor their personal or %conversations, family photographs, or 
confidential conversations and family photographs. When we suggested the option to disable the agent during personal conversations, participants considered it infeasible % to draw such boundaries 
in practice.
\begin{quote}
    ``\textit{I don't know how that [disabling the bot] could work because it's not like you want to say something so you're like, ``Okay, let me go and disable the agent,'' and only after that you'll say it. You just go with the flow, right?}''
\end{quote}

\bheading{Manual Activation}
Privacy concerns in automatic strategies prompted a discussion about the manual strategy. Participants suggested that the agent should refrain from actively monitoring the group. Instead, 
users could anonymously invoke the agent if they perceive a message as harmful. The agent would then get activated and start the anonymous deliberation process within the group. Even though the agent would still exist in the group, participants found comfort in thinking the agent would not ``scan'' all messages in the group.

Apart from safeguarding privacy, participants felt that, unlike the automatic techniques, manual activation would give them the agency to activate the agent and decide when it should be invoked and on what content. As P6 explained, he didn't want every harmful conversation to be flagged by the agent:
\begin{quote}
    ``\textit{This would be an excellent strategy, because, for instance, the group that I have with my close friends...we are accepting of different political and individual ideologies of our friends. Normally we don't talk about problematic or offensive things. We just normally share jokes or gossip.}'' -- P6
\end{quote}
%As an outcome of this agency, 
Participants pointed out that this way, the agent would be activated less frequently and save them the time and effort needed to respond to the agent. %Finally, this strategy allowed participants to anonymously flag content by activating the agent. 
Moreover, they felt that invoking the agent anonymously would enable them to safely express
%thought, was a benefit in itself, as it provided them with a safe way to express 
dissent in the group and nudge group members to not share problematic content.

% Participants also agreed that all group members---not just the admins---should be allowed to activate the agent in response to an inappropriate message.
%Despite these benefits, this strategy posed the risks of upsetting group dynamics in smaller, close-knit groups, such as friends and family. 
However, participants believed that in smaller, close-knit groups, where members knew each others' opinions well, people could guess who might have activated the agent even if it was done anonymously. This concern raised the potential for accusations and conflicts within the group, negatively impacting group dynamics. Hence, participants unanimously agreed that manual activation would be more suitable in larger groups.

% tension between privacy and detecting harmful content
Overall, these findings show the inherent trade-off between privacy and fostering positive group dynamics. While automatic approaches may compromise privacy, they have limited impact on disrupting group dynamics (since group members cannot be accused of setting off the bot). Conversely, manual approaches would preserve privacy but might hurt group dynamics. \label{design_tension:privacy}

\subsubsection{Participation: Who Should Deliberate?}
% Using an agent to facilitate deliberation requires other design decisions. For example, who should deliberate, how long should the process last, and what if no one is interested?
% After initiating the deliberation, the agent would have to decide on who should deliberate and for how long. Below, we explore participants' views on these questions.

% \bheading{Who should deliberate?}
The probe showed that after getting activated, the agent would reach out to group members to solicit their opinions. We asked participants which group members the agent should reach out to. Two primary strategies emerged: reaching out to all members or a random subset of the group. These strategies had trade-offs in terms of representation and workload.

% Participants agreed that asking all members. However, the loud voices in the group would always get a chance to speak up and would drown out the voices of the minority. On the other hand, a random subset ran the risk of leaving out the minority opinion from a random sample, which would skew results. This results in a difficult situation where asking a random subset could leave out the minority opinion, but asking everyone could drown out minority voices, leaving a potential under-representation of the minority in both cases.
Participants agreed that soliciting opinions from all group members would allow everyone (including gender, political, or other minorities within the group) to express their views. However, they also recognized that if everyone's opinions were taken, %then 
the majority viewpoints would dominate each conversation, overshadowing the perspectives of the minority.
\begin{quote}
    ``\textit{If it's asking everyone, then some people will just be able to answer every time.
    % But if it's few members, then you have a greater chance that it would reach out to parties who might have an independent opinion.
    }'' -- P6
\end{quote}
Hence, P6 argued to invite a random subset of group members for deliberation. However, other participants raised concerns that a randomly selected sample would risk excluding minority opinions, leading to skewed representation and outcomes. This dilemma highlights the difficulty participants reported in balancing the inclusion of minority opinions, as both approaches---asking everyone or selecting a random subset---could result in an unfair representation of minority viewpoints. %voices.

Another common concern among the participants was that participating in the deliberation would increase their workload. % Moreover, the deliberation would force them to engage with harmful content which sometimes they otherwise can ignore. 
With the first strategy (asking all members), participants were concerned that the agent would reach out to them too frequently, which might turn them off. Further, in larger groups, this strategy would lead to a summary message with a long list of opinions that would be arduous to read. Therefore, participants suggested asking for a random subset of group members. However, that ran the risk of not receiving enough opinions in smaller groups. Considering these trade-offs, P10 suggested a hybrid approach: asking all group members for their opinions in smaller groups (up to 10 members) and asking a small random subset of the group in larger groups. P10 elaborated: 
\begin{quote}
    ``\textit{So, maybe...you do a cap of 10 [i.e., asking 10 members for their opinions]. So in the smaller groups [with $\leq 10$ members], everyone gets reached out to and in bigger groups, 10 people get reached out to at random.}''% -- P10
\end{quote}

We also proposed two other strategies in the design worksheet based on prior work. The first option was based on crowdsourcing to identify harmful content~\cite{kazemi2022tiplines, Kim2018, Bhuiyan2020crowds} and suggested that the agent could reach out to people \emph{outside} the group for their opinions. %This was based on work that suggests crowdsourcing as a way . 
The second option involved asking only ``senior'' group members (e.g., elderly members, domain experts, people in positions of power) based on prior work that found the existence of credentialed ``gatewatchers'' in WhatsApp groups~\cite{2022accost}. To our surprise, our participants unanimously struck down both these ideas. They believed that people outside the group would not have the context of their group and that there was no way to measure ``seniority'' and ``expertise'' in WhatsApp groups. P6 shared: 
\begin{quote}
    \textit{``How do you determine senior members? Because WhatsApp doesn't necessarily ask you your age. And if seniority is determined by how long have you been a part of a group, that doesn't mean anything because people might just change their number and leave the group and that would affect their seniority. So I don't think seniority [of any kind] should have anything to do with the agent.''} %-- P6
\end{quote}

% Finally, our findings suggest that platform-level crowdsourcing may not be suitable for WhatsApp groups that are closed, end-to-end encrypted, and context-specific, unlike open platforms like Twitter~\cite{allen2021crowds, Bozarth2023RedditCrowds}. 

\subsubsection{Duration of the Deliberation}
We also asked participants how long the agent should wait for responses before %summarizing them and 
sending a summary back to the group. This was an important design decision because, as P4 said, ``\textit{You're demanding someone's time who might be busy.}''

Participants suggested that the agent could wait until everyone responded. However, they pointed out that some people may be inactive on WhatsApp and might never reply. Hence, participants proposed an alternative --- that the agent should wait for a predefined amount of time (a few minutes, hours, or days). % However, they agreed that two factors had to be balanced in deciding how long this should be: access and context. 
They said that if people are given too little time, they might not get enough time to read, think, and respond, particularly if the agent requests them while they are busy with work, personal chores, or sleeping. %, sleep, children, or household chores. 
On the other hand, if the agent waits too long, the group's context may change and conversations might move on, especially in larger groups. P8 described: 
\begin{quote}
    ``\textit{[If the agent waits too long,] people may start talking about something else completely... Maybe people are over it and it'll re-instigate the whole thing again... Like, if people are ready to get over it, there's no point of the agent to [force the deliberation].}'' % -- P8
\end{quote}
% Further, the longer the agent waits, the longer time people have to forward content without hearing the results.

% Initially, some participants suggested 2-6 hours, arguing that people check their phones every few hours. However, after more discussion, they acknowledged that merely seeing the agent's request was insufficient since people need dedicated free time to contemplate and respond. They also pointed out that the agent's request may come during inconvenient times, such as at night or during working hours. Hence, eventually, 

Hence, participants went back and forth in deciding the optimal waiting time, trying to balance the availability of group members and the context of the group. After discussion, most participants agreed that if the agent gave them one day to respond, it would accommodate their work schedules and differences in the time zones of the group members. %, and provide sufficient time for people to stay informed with daily news. 
When asked about the risks of resurfacing harmful messages from the previous day, participants commented that ``\textit{what matters is whether the harmful content is called out or not}'' %it was important to call out harmful content 
and that conversations would remain recallable within a day. %Further, if opinion requests accumulate during the day, this would give them time to address them in the evening. 
P14 described: 
\begin{quote}
    ``\textit{Group members should be given one day because... they can only check when they get some off time during the day. For instance, if somebody is doing a 10-5 job then after coming back at 5, they will check the group [and the agent] in some time, after they have relaxed.}''% -- P14
\end{quote}

One participant, P2, also suggested waiting for a certain number of responses, rather than a predefined time, or a combination of the two: ``\textit{If a certain fraction of people reply, the agent should just report back. If not, maybe after a certain time it should report back.}'' However, P11 said that ending the process early would open the doors for bias, as 
vehement defenders would respond quickly, fill the spots, and leave no room for opposing views. This shows that seemingly unrelated design decisions such as the duration may skew the deliberation process and shape whose voices get included. %in the deliberation process.

\subsection{The Process: Anonymous Deliberation} \label{subsec:anon_deliberation}
Next, we move on to the second dimension of deliberation: the process, which includes the actions of the participants and the quality of their communication. % We now discuss the preferences of the participants about how they would engage in the deliberation process facilitated by the agent. But first, we must clarify the peculiar nature of deliberation in this context.
Below, we explain how participants envisioned participating in the deliberation process facilitated by the agent.
% collects opinions, collates them, and sends them back to the group is the deliberation process itself. It might be assumed that the ``deliberation'' is a free-flowing discussion that follows after the summary is sent to the group. However, close ties in WhatsApp groups prevent such open discussions. Hence, sharing their views with the agent---which then posts them anonymously---is the deliberation itself.

\subsubsection{Responding to the agent}
% The first aspect of deliberation shown in the probe was when participants would share their opinions about a message with the agent.
Overall, participants preferred doing careful research before responding to the agent instead of instinctively responding to it, but were wary of the effort this might require. As a result, they said that instead of responding to the agent always, they would respond selectively.

\bheading{Researching Before Responding}
In our probe, when the agent would reach out to participants to solicit their opinions, it would ask them to rate the message and provide reasoning behind their rating (see Figure~\ref{fig:bot_design}b). To our surprise, very few participants said they would respond to the agent based on their instincts or other indicators (e.g., emojis, bold, sender's reputation, etc.). Instead, most participants preferred researching the topic by looking it up online or asking friends before responding to the agent. They believed that it was important to know all the facts before presenting their opinion. P14 mentioned: % Hence if they didn't know enough about the topic already, they would research it before responding to the agent.
\begin{quote}
    ``\textit{How can we just make a statement without actually knowing in depth about it... But if the agent asks me something that I know about already, I will answer immediately.}'' %-- P14
\end{quote}

% This was a surprising result as we expected that most participants would respond to the agent based on their instincts or other indicators (e.g., emojis, bold, sender's reputation, etc.). However, very few participants said so; most said that they would research thoroughly before responding. Indeed, one participant initially said that she would assess the content based on her gut feeling. However, she later realized that relying solely on her intuition wouldn't allow her to give a confident response to the agent. Hence, when the agent would ask her to rate content, she would do a quick Google search to find out more about it and rate it an absolute zero or ten, instead of something in between.

\bheading{Too Much Effort}
Participants expressed concern that anonymous deliberation would increase their workload. In order to reply to the agent, they would now have to pay attention to the content that they would have otherwise ignored. Similarly, responding to the agent would require research and typing out an appropriate response, which would mean more effort. Participants believed that doing this additional work on something they otherwise could ignore might cause them more stress. They mentioned that group members might put in this effort initially, but over time, they would respond briefly or not respond at all. This theme of additional workload was recurring in our study (also seen in the previous section), highlighting a design tension between combating harmful content and minimizing user workload. \label{design_tension:burden}

Participants suggested some approaches to reduce their workload in researching content and responding to the agent. For example, they asked if the agent could help them find credible sources of information relevant to the current content, instead of having to search themselves. P10 emphasized: 
\begin{quote}
    ``\textit{Cut out [i.e., automate] the verification part, the part that my grandmother can't do by herself. If that was done for her...[that would be better].}'' %-- P10
\end{quote}
Further, a few participants wanted to reduce their effort in typing out their opinions. They believed that there would be a finite set of reasons why a certain message would be harmful; the agent could provide them with options (e.g., drop-down list, WhatsApp polls) to choose from instead of requiring them to type their opinions every time. They also suggested that the agent could help them write reasoning that presents constructive arguments in an emphatic tone.

% Participants also suggested that the agent could send notifications and requests when they are available (e.g., evening hours) to limit distractions. 

Participants also pointed out the sheer volume of harmful content they received across multiple groups and were worried about getting spammed with too many requests from the agent. They envisioned finding it especially annoying or distracting if the agent reached out to them while they were working. As such, participants didn't want to respond to the agent more than 1-3 times a day and hence expected to sometimes ignore the agent. They also suggested that the agent could pool requests and send notifications when they are available (e.g., evening hours) to minimize distractions.

\bheading{Selective Responding}
One strategy participants suggested to reduce workload was to be selective in responding to the agent. They commented that they would respond only during ``peak events'' such as when societal tensions were heightened (e.g., riots) or when the message could affect big outcomes (e.g., elections). P10 distinguished between big and smaller peaks:
\begin{quote}
    ``\textit{It depends on the pulse of the nation... During COVID, I cared a lot about accessing the credibility of health information because the infection was potentially life-threatening. Right now that movie came out, Kerala Stories\footnote{A religiously divisive film that was touted to have won an Oscar when in reality it was only India's entry for the award.}, and there have been messages about that. That's like a small peak, so if I had the agent right now, it would be helpful...}''% -- P10
\end{quote}

Similarly, other participants only cared about debunking content in a few groups (e.g., family groups) and stated that they would ignore the agent's request if the message was received in other groups (e.g., neighborhood groups). Another set of participants said that they would reply to the agent only if they were interested in the topic. P11 stated: 
\begin{quote}
    ``\textit{I think it depends on how much the topic affects you. Like if it's something very personal to you, then you might get quite triggered by it and be like, yes, I have to prove them wrong. But if it's just something you find stupid, you're like, okay, these people are foolish to believe it, but you don't really care to change their opinion.}'' %-- P11
\end{quote}
Instead of receiving such requests from the agent and having to ignore them, some participants preferred the option to opt out of the agent's requests in groups where they were not active or in topics they were not interested in addressing.

\subsubsection{Perceptions of the Summary Message}
Our probe showed that after collecting opinions from group members, the agent would compile them into a summary message to be sent back to the group. While participants appreciated reading other group members' opinions, they emphasized that the summary was only a collection of opinions and should not be treated as a fact-check on the shared message.
% Participants found this summary message helpful in understanding others' opinions but emphasized that it only surfaces opinions, not the ground truth.
% They appreciated anonymity but had concerns about de-anonymization and the length of the summary message, and recommended AI capabilities to improve it.

\bheading{Anonymity}
All participants agreed that the opinions presented in the summary message should be anonymous, i.e., they should not be attached to the names of the group members. They believed that anonymity would help them speak up without worrying about repercussions when harmful content is shared by those in a position of power (e.g., boss, school principal, elderly family member), a concern also raised in prior work~\cite{2022accost}. P13 elaborated: 
\begin{quote}
    ``\textit{No, the names should not be mentioned. Because if something good happens, people will be happy about it, but if it is the other way around, they will take offense to that.}'' %-- P13
\end{quote}

One participant, P10, preferred staying anonymous most of the time but not always, so she suggested making it the default but optional. She said:
\begin{quote}
    ``\textit{I would identify myself in situations where I have expertise and clout, where people might be inclined to trust me. Where they won't trust me, I won't identify myself.}'' % -- P10
\end{quote}

Participants expressed that anonymity made the agent non-confrontational, that this was its ``biggest'' strength. Given this unanimous preference for anonymity, we asked participants if any features of the summary message might lead to de-anonymization. In the summary message shown in the design probe, members' opinions were displayed in serial order of when they were received. However, participants felt that the opinions should be displayed in a randomized order. They argued that some people are chronically online, so it may be possible to identify people based on their response speed. Similarly, other participants thought that vehement opposers (or defenders) of the content would respond quickly, which could cluster similar responses at the top of the list and bias readers. So, they recommended randomizing the order to maintain neutrality.

Similarly, many participants noticed that it was possible to identify people, especially close ties, based on how they composed their messages, e.g., by their use of punctuation and shorthand. They asked if the agent could rephrase the opinions so that linguistic patterns could not be used to identify the person. They suggested that the agent could fix grammar, edit abusive or attacking words, and remove emojis to protect anonymity. However, some participants were opposed to the agent rephrasing their opinions and preferred to see a verbatim copy. They feared that the agent might misrepresent their feedback and spoil the \textit{``sanctity''} of user's opinions. Further, a few participants felt that people would value human opinions more and rephrasing by the agent would diminish the significance of the feedback. Some participants felt that people themselves might write differently to avoid being identified.
% Design tension between anonymity and misrepresentation. Shows that anonymity is hard in close private group chats.

\bheading{Intelligent Summarization}
%As discussed above, maintaining anonymity in the summary message was hard, hence participants suggested rephrasing opinions. However, one participant felt that even rephrasing wouldn't guarantee anonymity as people could still be identified by the opinion itself. 
While some participants were opposed to editing group members' opinions, many participants saw value in summarizing opinions into a single paragraph instead of listing each opinion separately. Participants felt that doing so would not only safeguard the anonymity that participants dearly valued, but would also improve readability.
% and set the right tone for deliberation.
These concerns emerged especially for large groups in which participants felt that listing each opinion would lead to a long summary message.
% and people might use less civil language when expressing their opinions on divisive topics.
P7 expressed: 
\begin{quote}
    ``\textit{There needs to be some sort of summarization I'd imagine, because if it's a larger group... I wouldn't sit and read 30 different points from people or 50 different or 100 different points, which may have gone out.}'' %-- P7
\end{quote}

This way, participants believed that the agent could highlight the ``\textit{main points}'' by putting together the cumulative knowledge gained from different people. This would also reduce the workload as they would no longer ``\textit{have to write everything perfectly.}'' Further, in creating the summary, the agent could parse each opinion and prioritize those backed by evidence from credible sources, filtering out those that are uncivil, divisive, or not well-reasoned. 
% or ``wrong'' (participants assumed that the agent already knew what was wrong).

Some participants did not want to read even a textual summary and recommended that the agent produce visual summaries. They highlighted that they would only read the average rating reported by the agent without paying much attention to the reasoning. %. However, there was no consensus on this as other participants appreciated reading others' opinions. Another participant 
They suggested visual summaries where the agent could mark a positive, neutral, or negative opinion with red, yellow, and green, so that people could understand the prevailing sentiment with a quick glance.

\subsubsection{Reflection After Reading the Summary Message}
We also asked participants if the agent should do anything after posting the summary message such as explicitly encourage further deliberation or moderate deliberated messages.

\bheading{Encouraging further discussion}
Most participants agreed that the agent should only report the summary message in the group without explicitly encouraging further deliberation. They wanted the agent to let \textit{``the feedback do its own job''} and allow people to reflect and decide for themselves what they want to believe. This aligns with other decentralized approaches where users are empowered to form their own beliefs about the content they encounter online~\cite{jahanbakhsh2023personalized, Jahanbakhsh2022Peer}. Additionally, participants believed that individuals would naturally engage in further discussion if they had additional thoughts, without the agent needing to prompt them explicitly. P10 stated: 
\begin{quote}
    ``\textit{I don't think I'll read `you should discuss this' and then choose to discuss it.}'' %-- P10.
\end{quote}

Another reason for this choice was that any further discussion in the group would not be anonymous and could lead to bitterness. To avoid such conflicts, some participants floated the idea of a follow-up deliberation session facilitated by the agent itself. For example, they proposed that a few hours after posting the summary message, the agent could reach out to group members again and get feedback on whether their opinions have changed or if a stronger consensus could be formed given the opinions presented in the summary message. %However, this was only an initial idea and could not be developed further during our interviews.
% This would help gauge if people's opinions changed after reading the information in the summary message.

\bheading{Moderating deliberated messages}
We also asked participants if the agent should moderate messages that had received poor accuracy scores during deliberation. Although WhatsApp currently does not allow autonomous agents to moderate content, we wanted to explore the capabilities participants would like the agent to possess. 
The participants were ambivalent about moderating content based on the summary message, highlighting that the summary surfaced only opinions, not facts. For example, when we asked if the agent should delete a message from the chat if it received negative feedback, participants commented that actions such as deleting content or warning users might create a negative atmosphere. Instead, they recommended softer moderation strategies, such as disabling forwarding on messages while they are under deliberation (but not after). This balanced users' agency and safety: it would allow them to forward a message if they found it trustworthy even after reading the deliberation summary. P5 described: 
%However, participants were ambivalent about moderating content based on the summary message, highlighting that the summary surfaced only opinions, not facts. %Hence, they were conflicted about whether the agent should take ``invasive'' moderation actions based on mere opinions. In response, participants suggested disabling forwarding on a message \textit{until} the results of the deliberation were posted, but not after.
\begin{quote}
    ``\textit{Once a discussion happens and the results come, people will be more sensitized towards it... People who [would have] forwarded it mindlessly will [now] be more wary of forwarding if three people are not agreeing to it. And the discussion might lead to somebody going on the Internet and finding out a more credible source. That will stop many people from forwarding the message.}''% -- P5
\end{quote}

Another approach that participants recommended was to tag messages that received low scores. For example, participants mentioned that the agent could make visual changes, for example, change the font color to grey to reduce emphasis for future readers or add a red exclamation mark to emphasize the questionable veracity. They believed that doing so would reduce the illusory truth effect~\cite{2022usenix_students} where users start to trust harmful content the more they see it. However, participants realized that tagging came with the trade-off that it would not stop unconvinced group members from forwarding the content. These findings highlight a design tension between maintaining user agency in sharing content and stopping the circulation of questionable content. \label{design_tension:agency}

\subsection{Outcome: Is Anonymous Deliberation Helpful?} \label{subsec:is_anon_delib_helpful}
\citet{Friess2015} define the \textit{outcome} as the expected results of the deliberation. Throughout our study, we were careful to dissociate the approach (i.e., anonymous deliberation facilitated by an agent) from its specific instantiation (i.e., our design probe). This was important to understand the impact of anonymous deliberation in combating harmful content. Participants noted both the strengths and pitfalls of using conversational agents to facilitate anonymous deliberation. 
% They raised valid questions: How do we know whether people's views change with deliberation? Will deliberation surface the truth or only the majority's opinion? We describe these findings below.

\subsubsection{Strengths of Anonymous Deliberation}
Most participants agreed that the agent would help identify and contest harmful content in their groups. They felt that the agent would help \textit{``filter the right from the wrong''} using knowledge within the group, and help them hear each others' opinions. They described several benefits of anonymous deliberation.

\bheading{Anonymity}
Most participants said that the anonymous nature of the deliberation would allow them to speak out against harmful content. Most of the participants were part of WhatsApp groups in which they maintained offline social ties with other members of the group. Hence, they feared that deliberation without anonymity would jeopardize their relationships with group members or upset authoritative figures, such as elder family members, superiors at work, among others. % as also found by \citet{2022usenix_students}.
% For this reason, participants would not always express their opinions about the content posted in their groups.
%So, they appreciated the non-confrontational nature of the agent.
As a concrete example, P15 described how such an agent can help parents express displeasure in a WhatsApp group in which her child's school principal often sends political propaganda, but no parent has been able to speak up for fear of backlash for their children. P2 elaborated on how anonymous deliberation might help address power differentials in groups and help surface more reasoned voices: 
\begin{quote}
    ``\textit{The most extremist voices are the loudest, and that does not represent the group. There are enough people who believe the other way, and just knowing that fact can help.}'' %-- P2
\end{quote}

Some participants emphasized that anonymous deliberation through an agent would help limit long-drawn arguments and encourage group members to feel more comfortable speaking up. For example, P11 felt that anonymity would limit personal responsibility: ``\textit{If the person is not convinced, you don't feel personally responsible to continue defending.}'' %she said. This would limit long-drawn-out arguments, and more group members may begin to feel more comfortable speaking up.

\bheading{Neutrality}
Participants emphasized that any agent that facilitates deliberation must be neutral and preserve users' agency. For example, they appreciated that the agent in our design probe asked for opinions instead of telling them what was right or wrong, making them feel respected and important. % They explicitly pointed out the neutrality of the agent design, which we describe below.
They felt that since the agent would be a neutral mediator in the group, it would encourage people to focus on the facts and avoid personal attacks. While negative feedback on content could make the sender feel attacked, participants felt that mediation through an agent would dilute such feelings compared to direct confrontation with other group members. P10 explained:
\begin{quote}
    ``\textit{If I say, `Uncle you are wrong,' that's an issue. But if the agent states that a member said this might be false, that comes across as mild and acceptable.}'' %-- P10
\end{quote}
%Hence, the sender wouldn't feel attacked upon receiving other's opinions. 

Participants also felt that a neutral agent would help impartial group members form an opinion without alienating them. For example, P6 explained that when arguing about a message, the discussion tends to get heated and people use less polite language. Such heated conversations drive away people who are yet to make up their minds as they tend to focus on the argument's tone rather than the facts presented. He believed that a neutral facilitator would encourage people to comment on the veracity of the content instead of expressing their political opinions. He said:
\begin{quote}
    ``\textit{An agent which is not politically inclined gives context that it's not about different political opinions, it's about the truth, it's about decency, it's about morality...}'' %-- P6
\end{quote}

\bheading{Accountability}
Participants felt that not only will deliberation help people identify harmful content propagating in their groups, it will also encourage them to \textit{``think twice''} before sending suspicious or misleading messages to avoid being flagged by the agent. Hence, P4 believed that the deliberation process would create accountability within group members.
%She suggested that this preventative effect was due to a social connotation of appearing naive or uninformed to other group members upon being flagged by the agent. 
P3 elaborated:
\begin{quote}
    ``\textit{It's a common tendency of human psychology that whenever they see negative information, without even giving it a thought, they circulate it to a large number of people... The agent will stop people from forwarding the messages blindly... What the agent does is at least it gives people a second thought and [a chance] to develop interest around that information...}'' %-- P3
\end{quote}

%In the same vein, P4 believed that the deliberation process would create accountability within group members. Since the agent would call out harmful content, group members would become conscious of sharing questionable content, which would slow down its spread.
% of knowing that one is accountable to other group members for sharing veracious content

\bheading{Strength in Numbers and Diverse Opinions}
A strong argument for deliberation was the strength in numbers. Participants emphasized that the groups they inhabited had diverse people who likely knew whether something was harmful or not, but hesitated to speak out to avoid conflicts with other group members. For example, P13 mentioned that her groups have journalists, advocates, police officers, media people, social workers, and doctors, who had opinions they felt uncomfortable sharing. Participants felt that the deliberation process would help surface the opinions of such group members. They drew a parallel between the agent and the \textit{panchayats}\footnote{A panchayat comprises elected representatives in Indian villages who collectively make decisions on local governance and conflict resolution.}, highlighting that both processes are based on decision-making through deliberation. P14 articulated:
\begin{quote}
    ``\textit{In a group of ten people, all of them are never going to label something accurate if it is not...And if all ten consider it accurate, then there must be some authenticity to it.}''% -- P14
\end{quote}
% Indeed, one participant called this a group-level crowd-sourced fact-checking process to improve the context around potentially harmful content on WhatsApp.

% Further, these participants believed fundamentally in the power of diverse opinions.
Participants emphasized the importance of hearing diverse opinions. They believed that a ``\textit{collective thought process was better than everyone's individual opinions}'' (P20). They emphasized that hearing everyone's opinions would enable them to understand other perspectives, which would either help them recognize misunderstandings, confront their own biases, or, if they're right, find more support for their worldview.  P6 elaborated: 
%Simultaneously, they believed that hearing others' opinions would provide support for their stance by identifying individuals who shared their perception of the content as harmful.:
\begin{quote}
    ``\textit{In big groups, there is necessarily some kind of diversity. People of different political inclinations are members of the group. So, if there is a problematic news article shared by a person and I find it offensive...then I may receive the deliberation from others who are not of the same political inclination as me but they also perhaps find it problematic, right? So it would just increase the veracity of the deliberation, that the deliberation is not propelled by one group. The deliberation itself is diverse...All sorts of people have problems with this message. And then, a neutral party should be aware that this message perhaps is not in the best of their interests...}'' 
\end{quote}

\subsubsection{Pitfalls of Anonymous Deliberation} \label{subsubsec:pitfalls}
While acknowledging the strengths mentioned earlier, participants also pointed out the drawbacks of engaging in anonymous deliberation through a conversational agent. In addition to privacy concerns about the content that the agent would have access to, they were worried about the impact on group dynamics and limited usefulness in certain contexts.

\bheading{Impacts Group Dynamics}
Many participants said that the deliberation process could upset group members. The sender could feel attacked by the agent getting activated or by receiving unfavorable feedback from others. In small groups of friends and family in which people have strong social ties and know each other's ideologies or linguistic styles, participants worried that it would be easy to identify who said what, which might lead to bitterness, accusations, and conflicts.  They were also concerned that some group members may misuse the veil of anonymity to make disrespectful comments and personal attacks. In line with the amplification theory proposed by \citet{toyama2011amplification}, participants emphasized that an agent can only amplify the group's existing intent to contain harmful content, not replace it. P11 said:
\begin{quote}
    ``\textit{But I think it ultimately comes down to how the discussions in general go in the group. If it's a little hostile environment in general, it is going to be headed that way, you can't really avoid it.}'' %-- P11
\end{quote}

Going beyond groups with close ties, participants pointed out that the agent may lead to fights even in large public groups where people may not know each other or their ideologies. P19 compared such groups to public social media platforms:
\begin{quote}
    ``\textit{Anonymity doesn't matter... In large groups, in YouTube comments, people don't know each other, so it's sort of anonymous, but fights still break out. It's about whether something was said that's against their opinion.}'' %-- P19
\end{quote}

\begin{figure}[t!]
    \centering
    \includegraphics[width=0.9\textwidth]{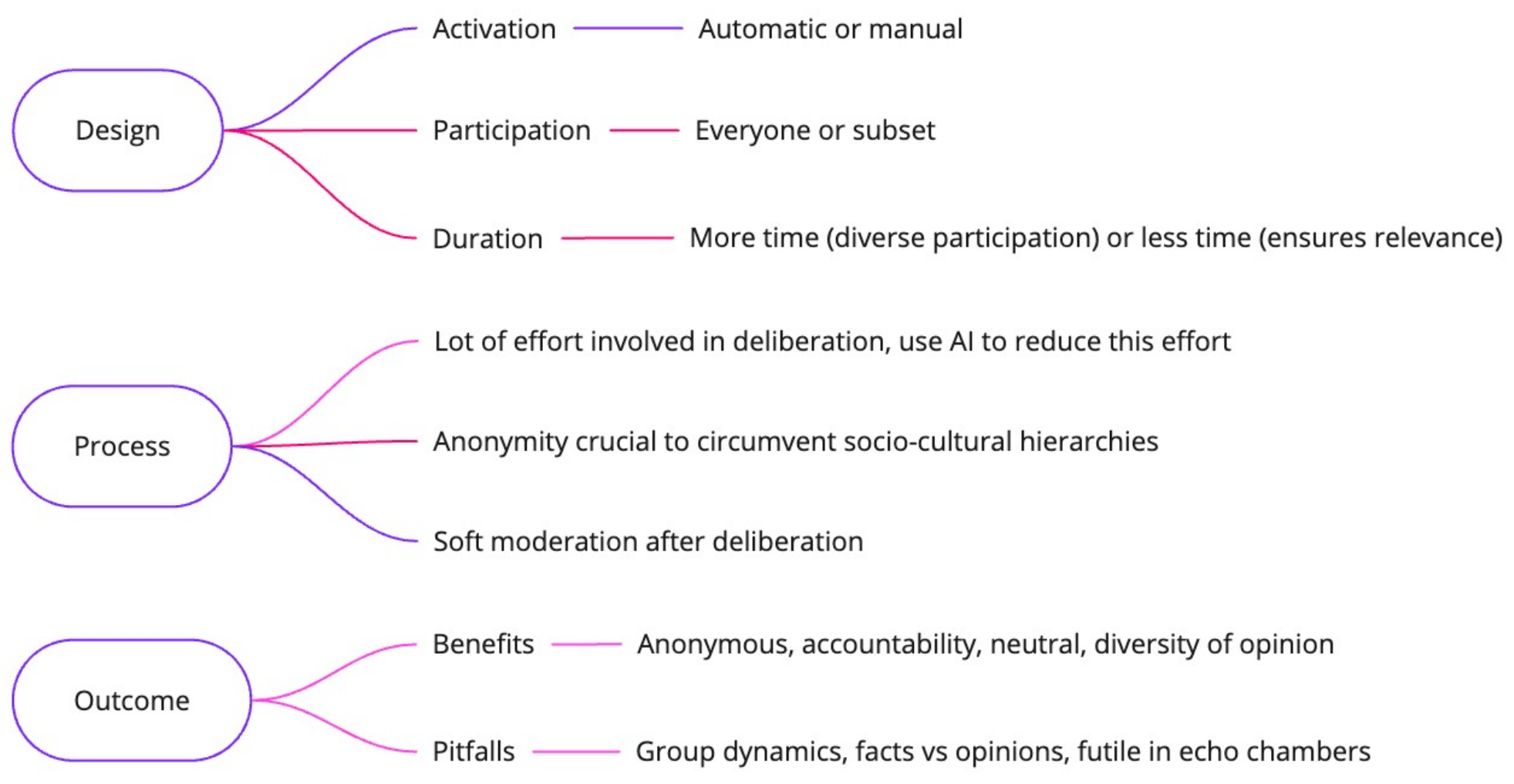}
    \caption{Summary of findings from our study}
    \label{fig:findings}
\end{figure}

\bheading{Flagging More Helpful than Deliberation}
Some participants worried that the deliberation process would surface opinions---which could be biased---instead of facts. Hence, they wouldn't trust the process to debunk fake news or misinformation. However, they found the agent useful for \emph{flagging} harmful content. They believed that an indication of a message being potentially harmful was more useful than the deliberation itself. They felt that a message being flagged would encourage people to do their own research and reflect on its contents. P14 stated: 
% For example, seeing people discuss a message may inform a neutral onlooker that it may be of questionable authenticity.
\begin{quote}
    ``\textit{If the agent gets activated, it tells me that there may be something fishy going on and then I will go and search for it.}'' %-- P14
\end{quote}
In this sense, the flagging of content by the agent can be construed as accuracy nudges, which have been shown to improve sharing attitudes~\cite{2021accuracy_nudges}. Another participant liked how Instagram and Twitter would similarly flag posts containing potentially harmful content. P7 articulated: 
\begin{quote}
    ``\textit{The approach that everybody [other platforms] seems to be taking online is that here's some questionable content, so I will write `questionable content' in bold letters below it. Now... I am less likely to forward it because then whoever gets it will see that I sent them something that's `questionable content'. So there's some social reputation or social currency which is at risk...}'' %-- P7
\end{quote}

%We previously talked about another social benefit of the deliberation process: that people might start to worry about setting off the agent and therefore ``think twice'' before sending unreliable content. In fact, even this worry is associated with the flagging of their message, not with the deliberation process that follows. Hence, 
Participants felt that the deliberation process was a very roundabout way of achieving the benefits that could be achieved simply by flagging harmful content. They argued that 
% He suggested that instead of following a long-drawn-out process and still only achieving the benefits of flagging content, WhatsApp could simply just flag questionable content.
WhatsApp could simply show a popup asking users to confirm if they wanted to forward a message to others, a strategy also recommended by \citet{2022usenix_students}. However, some participants felt that simply flagging content would lose value over time as people would ignore the flag, much like how people ignore the labels (``Forwarded'' and ``Forwarded many times'') used by WhatsApp~\cite{Hall2023Beyond}. One participant emphasized that deliberation is better than flagging as it helps surface \textit{why} something is questionable and not just \textit{whether} it is questionable.

\bheading{Futile for Echo Chambers}
Some participants believed that deliberation would not be helpful when people have strong opinions. They did not expect the quality of content to improve in extremist political or religious groups that \textit{``relish in sharing problematic content.''} They emphasized that deliberation was only helpful in moderate groups where some individuals were neutral. P7 mentioned: 
\begin{quote}
    ``\textit{If you're talking about the extremes of hate speech, things which lead to lynchings and stuff, I feel a majority of that information is in groups which are echo chambers... There is a bell curve of people, where there are some people on the fringes and a majority of them are in the middle and those are the people you're targeting this agent for.}'' %-- P7
\end{quote}

Similarly, participants felt that deliberation would not prevent harmful content from users who
% share such content but may not realize that it's harmful. These users
may genuinely believe in the information they encounter on WhatsApp. P7 gave one such example:
\begin{quote}
    ``\textit{This person probably thinks, `Oh my god, vaccines are going to kill us... I'm genuinely concerned for my friends and family, so I need to share this information with them.' In that case, I don't believe they're going to not post it.}'' %-- P7
\end{quote}
While the agent would flag such content and invite deliberation upon it, P7 was doubtful that it would teach naive users how to identify such content in the future or discourage them from posting such content. For this reason, some participants advocated building media literacy through conversational agents rather than countering each specific instance of potentially harmful content.

\section{Discussion} \label{sec:discussion}

%Our findings surfaced several tensions in designing a conversational agent to facilitate the deliberation of harmful content on WhatsApp groups. We also discussed how participants would interact with such an agent and the perceived benefits and pitfalls of deliberation to combat harmful content in WhatsApp groups. A recurring limitation of this approach was that it only surfaces individual opinions, not facts. Participants were concerned that individual opinions may be biased or misinformed, which may lead to incorrect conclusions from the deliberation process. Further, deliberation was expected to be futile in changing the behavior of users with extreme views, who have the potential to cause the greatest social harm such as riots. 

\begin{table}[t!]
\smaller
\centering
\begin{tabular}{|ll|l|l|}
\hline
\multicolumn{2}{|c|}{\textbf{Design Tension}} & \multicolumn{1}{c|}{\textbf{Description}} & \multicolumn{1}{c|}{\textbf{Findings}} \\ \hline
\multicolumn{1}{|l|}{Privacy} & Group dynamics & \begin{tabular}[c]{@{}l@{}}Is it acceptable to risk privacy in\\ order to maintain group harmony?\end{tabular} & \ref{design_tension:privacy} \\ \hline
\multicolumn{1}{|l|}{Freedom of speech} & \multirow{3}{*}{\begin{tabular}[c]{@{}l@{}}\\Circulation of\\ harmful content\end{tabular}} & \begin{tabular}[c]{@{}l@{}}Should an agent call out hate\\ speech in informal conversations?\end{tabular} & \ref{design_tension:freedom_of_speech} \\ \cline{1-1} \cline{3-4} 
\multicolumn{1}{|l|}{Burden on user} &  & \begin{tabular}[c]{@{}l@{}}Should users have to put in effort\\ to combat harmful content online?\end{tabular} & \ref{design_tension:burden} \\ \cline{1-1} \cline{3-4} 
\multicolumn{1}{|l|}{User agency} &  & \begin{tabular}[c]{@{}l@{}}Should users be allowed to forward\\ content known to be harmful?\end{tabular} & \ref{design_tension:agency} \\ \hline
\end{tabular}
\caption{Design tensions revealed in our study. The first two columns show the dimensions that may be in conflict while designing systems to deliberate on harmful content.}
\label{tab:design_tensions}
\end{table}

Our findings are summarized in Figure~\ref{fig:findings}. They surfaced several strengths and limitations of using conversational agents to facilitate deliberation on harmful content in WhatsApp groups. They also revealed several design tensions, summarized in Table~\ref{tab:design_tensions}. Using deliberative theory as a lens, we now discuss how deliberation was actualized in our study and its efficacy in combating harmful content. % ways to improve deliberation in WhatsApp groups. 
%Below, we discuss these points about the efficacy of deliberation.
We then compare deliberation with other approaches to combat harmful content. Then, we distill implications for designing an effective deliberation system for combating harmful content in end-to-end encrypted online spaces like WhatsApp groups that are outside the oversight of platform-mediated moderation. Finally, we discuss the limitations of our study and directions for future work.

\subsection{Actualization of Deliberation}
Deliberation is a discussion between individuals to better understand an issue, form opinions, and inform decision-making~\cite{Beauchamp2019}. The mechanics of deliberation that emerged in our study context had some differences compared to conventional deliberation spaces imagined by scholars of deliberative theory~\cite{Habermas1985, Cohen1989, Fishkin2011}. These differences emerged primarily due to the participants' unwavering preference for anonymity in the deliberation process. For example, %First, our example agent encouraged participants to engage critically with the content by asking them to rate it and provide reasoning behind their rating. %Indeed, our participants said that they would research the message online or by asking friends before responding to the agent. 
%Second, 
the agent collated members' opinions into a summary message to allow them to read others' opinions, reflect on them, and set the stage for further deliberation. 
While our participants appreciated reading diverse opinions in the summary message, they were ambivalent about further free-flowing deliberation as that discussion would not be anonymous.

This actualization of deliberation was different from the traditional conventional process that would typically involve a back-and-forth among the discussants. Hence, the complex sociocultural context of WhatsApp groups manifested deviations from traditional deliberation processes, especially when the content being discussed could be divisive. That said, the deliberative process that emerged still adheres to the key principles of deliberative theory~\cite{Fishkin2013, Habermas1985}. The anonymous deliberation allowed \textit{public reasoning} by facilitating the exchange of ideas asynchronously. The anonymity afforded \textit{inclusivity} and \textit{equality} by allowing members lower in social hierarchies to present their views without fear of upsetting power differentials. Finally, the process encouraged \textit{critical reflection} by asking users to rate and reason about harmful content. Hence, while deliberation might actualize differently in this context, it still adheres to key properties of the deliberative process and holds promise in allowing individuals to inform opinions on harmful content propagating in WhatsApp groups.

\subsection{Efficacy of Deliberation}
Most participants in our study believed that anonymous deliberation was useful in improving the quality of discourse in their groups. They perceived many benefits of it, such as anonymity and neutrality, nudging people to improve sharing practices, and strength in the diversity of opinions. Participants also highlighted several limitations, including its futility in echo chambers.
% A recurring concern that participants emphasized was that anonymous deliberation only surfaces individual opinions and not facts. They were worried that individual opinions may be biased or misinformed, leading to incorrect conclusions from the deliberation process. Further, deliberation was expected to be futile in changing the behavior of users with extreme views, who have the potential to cause the greatest social harm, such as riots. While certain concerns, such as potential impacts on group dynamics, may be mitigated through improved design as we discuss in Section X, others
This concern about the \emph{efficacy} of the approach requires a deeper discussion. Our participants emphasized that anonymous deliberation surfaces individual opinions which may be biased or misinformed, and not facts. Hence, they were concerned about its efficacy in combating fact-based content such as fake news and misinformation. This raises several important questions: \textit{Is deliberation effective when it primarily surfaces opinions rather than facts? Why would people change their behavior without robust fact-checking, based on potentially conflicting opinions? Is there intrinsic value in the deliberation process even if it does not reveal facts?} We turn to deliberative theory to answer these questions.

Of the three components of deliberation (design, process, and outcome), early deliberative theory~\cite{Habermas1985, Cohen1989} focused disproportionately on the outcome. The goal of deliberation was to produce a rational discussion that necessarily led to a consensual outcome that all participants agreed upon. Soon, however, this ``consensual'' view of deliberation gave way to a ``procedural'' view: In the second wave of deliberative theory, the process itself, not the outcome, became the fundamental part of the deliberation. This resulted from a growing understanding that consensus is difficult or even impossible to achieve when interests are fundamentally opposed~\cite{Beauchamp2019}. In addition, focusing on the outcome was futile because a fake consensus was easy to reach without following a deliberative process at all, for example, by forcing people into agreement~\cite{Beauchamp2019}.

As a result, modern definitions of deliberation emphasize the process rather than the outcome. For example, \citet{Beauchamp2019} defines it as an extended conversation between people to gain a better understanding of some issue. These definitions are procedural instead of outcome-oriented. Shifting the focus from reaching a consensus to emphasizing the process enables deliberation within groups of self-interested actors~\cite{Mansbridge2010} or individuals with moral disagreements~\cite{Gutmann1998}, such as in WhatsApp groups where people may have conflicting opinions. Moreover, deliberation prompts individuals to confront problems, consider the context, and reflect on their own biases instead of overlooking them.

Thus, drawing on scholarly work on deliberative theory~\cite{Kim2021moderator, Beauchamp2019}, we argue that even without fact-checking or achieving consensus, the deliberation process brings value by fostering fairness, authenticity, diversity, and reasoned dialogue among participants. As ~\citet{Beauchamp2019} noted, it allows people to \textit{``learn facts, ideas, and the underlying conceptual structures relating between them.''} In addition to our findings, which provide empirical evidence in support of deliberation, other studies also note that deliberation motivates people to read materials that aid in individual learning and attitude change~\cite{Davies2012}. % Taken together, our findings and prior work on deliberative theory shows that 
Thus, deliberation becomes an end in itself, allowing individuals to gain knowledge and shape opinions within the complex landscape of WhatsApp groups.

\subsection{Deliberation, Moderation, and Fact-Checking}
% Expanding on the theoretical advantages of deliberation, we now discuss some of the practical benefits of using deliberation to combat harmful content in WhatsApp groups. The practical benefits become clear upon comparing it to popular alternative approaches. One such approach is moderation, which mainly deals with what should be done with harmful content \textit{after} it has been identified as such~\cite{jahanbakhsh2023personalized}. The predominant approach for the \textit{identification} of harmful content is fact-checking. Fact-checking has received significant attention with researchers studying fact-checking practices~\cite{juneja2022human, Micallef2022factchecking, haque2020factcheckers}, proposing AI-assisted fact-checking~\cite{Nguyen2018factchecking}, and users' reaction to fact-checks~\cite{Appling2022factchecking}.

Two main approaches have been identified by researchers to combat harmful content on social media: content moderation and fact-checking. However, both these approaches have proven ineffective on WhatsApp. Due to end-to-end encryption, platform-led moderation is infeasible in WhatsApp groups, while sociocultural factors deter group members and admins from moderating content~\cite{FarhanaNewPaper, Shahid2023decolonial}. On the other hand, since fact-checking is reliant on human effort, it is difficult to scale given the sheer amount of harmful content online~\cite{juneja2022human, Micallef2022factchecking}.
% Social media platforms use content moderation to reduce the propagation of problematic content. However, online spaces like WhatsApp groups with end-to-end encryption are outside the platform's oversight. As a result, the responsibility to identify, contest, and moderate content often lies with group members and admins of such online spaces~\mbox{\cite{FarhanaNewPaper}}. A growing body of HCI and CSCW work points to inequities and biases in content moderation~\mbox{\cite{Shahid2023decolonial}} and research shows that admins and group members on WhatsApp rarely signal their doubts and confront each other~\mbox{\cite{FarhanaNewPaper}}. These studies suggest that moderation has limited potential to combat problematic content traversing WhatsApp groups.

% Another predominant approach for combating problematic content is fact-checking which has received significant research attention. Scholars have studied fact-checking practices~\mbox{\cite{juneja2022human, Micallef2022factchecking, haque2020factcheckers}}, examined user engagement with fact-checks~\mbox{\cite{Appling2022factchecking}}, and proposed AI-assisted fact-checking~\mbox{\cite{Nguyen2018factchecking}}.
% However, since fact-checking is primarily driven by human effort, it is difficult to scale given the sheer amount of harmful content online~\mbox{\cite{juneja2022human, Micallef2022factchecking}}. % Given the sheer amount of harmful content online, it is important to aggregate potentially harmful content that must be fact-checked.

Deliberation can serve as a complementary approach to address the shortcomings of these existing infrastructures. For example, deliberation can help identify potentially divisive content that needs fact-checking; only the content unresolved through group deliberation can be sent to professional fact-checkers. Deliberation could also effectively address and contest hyperlocal misinformation and propaganda (e.g., news about the local community) that fact-checking organizations may not know about~\cite{seelam_fact-checks_2024}, but group members may have the collective knowledge to debunk. Similarly, deliberation amongst community members can serve as an effective tool to inform opinions on multimodal content which is hard to debunk automatically and unscalable to debunk by human fact-checkers~\cite{juneja2022human, Machado2019Brazil}.

Scholars have also highlighted challenges in disseminating fact-checked information~\cite{Micallef2022factchecking, juneja2022human}. A community-driven deliberative approach can increase the reach of fact-checks, as users exposed to fact-checks can share them with others during deliberation. This will not only increase the reach but also boost effectiveness, as receiving fact-checks from acquaintances has been proven more powerful~\cite{2022strongties}. Another limitation of fact-checking is language accessibility~\cite{juneja2022human, kazemi2022tiplines}. Since fact-checking is often manual and a centralized effort, it requires knowledge not only of the local context but also of languages~\cite{seelam_fact-checks_2024}, posing another challenge to scalability. In contrast, a decentralized approach where group members deliberate among themselves avoids this issue, as they are likely to share a common language.

\subsection{Implications}
Building on our empirical findings and theoretical underpinnings of deliberative theory, we now discuss the implications of our study on the \textit{design} and \textit{process} of deliberative systems tailored to tackle harmful content in WhatsApp groups. We omit discussion on the \textit{outcome}, as deliberative theory underscores the significance of deliberation regardless of its specific outcomes.
% we now present design recommendations for deliberation systems that enable combating harmful content in WhatsApp groups. Based on the framework from \mbox{\citet{Davies2012}} to design deliberative systems, we organize our recommendations into the following categories: population (\textit{who} deliberates), spatiotemporal distance (\textit{where} and \textit{when} will participants be while deliberating), communication medium (\textit{how} will participants communicate), and the process (\textit{what} will occur between participants). These categories emerged from a review of empirical research in online deliberative systems and adjacent fields such as CSCW, HCI, and computer-mediated communication~\mbox{\cite{Davies2012}}.

% \subsubsection{Purpose} In our context, the purpose of deliberation is to improve knowledge or opinions about harmful content in WhatsApp groups. \citet{Kim2021moderator} called for designing systems that are specifically tailored for the discussion of divisive and emotional topics such as harmful content. We responded to this call and found that the deliberation of contentious topics requires careful attention to anonymity, privacy, and group dynamics.

\subsubsection{Design of the Deliberative Environment} Our study underscores that a deliberation system must respect the private and intimate nature of WhatsApp groups. Participants emphasized the importance of anonymity in calling out harmful content. However, since instant messaging platforms like WhatsApp do not allow anonymous texting, there is a tension between what's needed to combat such content and what's possible on the platform. In such a scenario, anonymity can be maintained through asynchronous communication via an intermediary. A conversational agent can be this intermediary, facilitating deliberation in a forum-like and anonymous manner, even if instant messaging platforms may not provide these affordances by design.

While such an agent could work theoretically, it is still an open question if users will stop engaging with such an intermediary as the novelty wears off. To ensure continued adoption, designers must consider the practical challenges of deploying such an agent including the workload on group members and the potential for bias, particularly in underrepresenting minority opinions. To reduce the burden on group members, the agent must consider users' interests, expertise, and availability to discuss a topic.

In addition to ensuring continued adoption, designers of such systems must also strive to inspire effective discussion. The choice of communication medium can be crucial in this endeavor. Research has shown that people are more productive when speaking rather than writing due to the lower cognitive load~\cite{Gould1978}. Even within the context of harmful content, prior work found audio debunking messages more effective in correcting beliefs than text~\cite{2022strongties}. In addition, research shows that speech is more emotive and less hostile than text~\cite{Kraut1992task}. These studies provide a strong argument for using audio as a modality deliberating upon harmful content, which would be particularly useful in groups containing low-literate individuals and older adults who often struggle with typing~\cite{2018farmchat, Hagiya2016typing}. However, this choice becomes complicated in our context due to the requirements of anonymity since individuals in small groups can be easily identified by their voices. As also recommended by \citet{Davies2012}, future work could consider translating spoken words into text to reduce the cognitive load of deliberation while preserving anonymity.

\subsubsection{Deliberation Process}
% This category encourages designers to ask \textit{what} process will occur between participants to facilitate deliberation.
Our findings reveal the following suggestions for the deliberation process in the context of harmful content on WhatsApp. %Our study revealed the following implications about the process:
% is it \textit{facilitated}, what is the \textit{structure}, are participants \textit{identifiable}, and are they \textit{incentivized}~\cite{Davies2012}. Our recommendations emerge from these axes.

\bheading{Human-AI Collaboration} Past work has shown that both \textit{facilitation} and \textit{structure} improve deliberative quality and evenness of participation in online communities~\cite{Lee2020solution, Kim2021moderator}. Our study extended these findings to the context of anonymous deliberation on WhatsApp and showed that a facilitator agent can encourage participants to research content and hear others' opinions. In addition, our study surfaced new avenues for facilitation through human-AI collaboration. Instead of construing facilitation as mere logistical support (e.g., structuring the conversation, time management, encouraging participation), future deliberation systems could use AI to ease the deliberation process for individual participants. For example, an agent could help find reliable sources, intelligently summarize opinions, filter out personal attacks, and write constructive feedback. However, these benefits of AI must be weighed against issues of bias: which sources are ``reliable'', which opinions to leave out in the summary, what constitutes personal attacks, and what makes feedback constructive? Not doing so carefully may create accusations of ideological bent and cause distrust in the agent. However, the answers to these questions are complicated as they may change with the sociocultural context of the group. For example, the agent may need to filter out sarcasm in a culture that considers it offensive but not in one that considers it friendly. To circumvent these problems, future work could imagine AI agents as personalized deliberative partners that adhere to each user's personal opinions and biases. We discuss this next.

\bheading{Agents as Deliberative Partners}
Findings from our study and prior work~\cite{Davies2012} show that deliberation might benefit each individual more than the group as a whole. %~\cite{Davies2012}. Thus, it has the potential to improve individuals' knowledge and sharing behavior. 
Given these individual benefits, future efforts can focus on positioning a conversational agent as a partner in deliberation rather than merely facilitating logistics. In our probe, group members expressed their opinions to an agent, and group deliberation occurred when they read each others' opinions in the summary message. Instead of deflecting back to the group, the agent could hear members' opinions, pose counter-questions, and encourage deeper thinking. While individual reflection might not be useful to inform others' opinions, it would help users reflect on their own biases and worldviews. Self-reflection could be an effective way to achieve the same benefits achieved from group deliberation without risking group dynamics (e.g., in-fighting) and maintaining users' anonymity.

\bheading{Negative Effects of Anonymity} \citet{Kim2021moderator} recommended designing deliberative systems that respect interpersonal and social power dynamics. An emerging body of work shows that the existence of social ties in WhatsApp groups prevents deliberation of harmful content~\cite{2022strongties, 2022accost, 2022usenix_students}. Cultural norms also impact participation~\cite{Min2009east}, for example, confronting elders is looked down upon in certain cultures. Anonymity counters these effects and boosts both participation in~\cite{Davies2012} and quality of deliberation~\cite{Baltes2002}. Our study further provides empirical evidence that anonymity can improve deliberation when social dynamics are at play. However, anonymity may also increase \textit{``troublemaker''} behavior~\cite{tucey_online_2010} such as goofing off, hampering discussion, or making personal attacks, especially when discussing divisive content. Hence, an anonymous deliberation system must also introduce appropriate moderation strategies such as removing or rephrasing opinions with expletives or involving group admins to ensure the civility of the discussion.

\subsection{Limitations and Future Work}
We used a design probe to explore the utility of conversational agents to facilitate the deliberation of harmful content in WhatsApp groups. There are limitations to this approach. First, while our probe helped participants conceptualize the design, we acknowledge that they did not have the opportunity to actively engage with or experience the features. Consequently, our findings are based on participants' anticipated behaviors and responses rather than their actual interactions with the system. % Second, we have presented a theoretical grounding for the utility of deliberation in improving individual knowledge about harmful content. The lack of a real-world implementation means that we lack quantitative data to substantiate this claim.
Our decision to conduct an exploratory design study was equal parts ethical and practical. We were wary of deploying technology in private sociotechnical spaces in the absence of evidence of its efficacy. Hence, our study aimed at exploring the design space first. Having done so, a promising avenue for future research involves a Wizard-of-Oz study within real WhatsApp groups. This would enable a comprehensive evaluation of the agent using a variety of metrics devised to measure the efficacy of the design, process, and outcome of deliberation~\cite{Friess2015, Beauchamp2019}.

\section{Conclusion} \label{sec:conclusion}

We presented a qualitative study to explore the design of a conversational agent to facilitate the deliberation on harmful content in WhatsApp groups. Using a design probe, we asked WhatsApp users in India about the strengths and limitations of using deliberation to combat harmful content, their interaction with such an agent, and issues related to trust, privacy, and group dynamics. Participants found this an effective approach to combat harmful content by hearing diverse perspectives, though they perceived a potential risk to group dynamics. They envisioned researching the content and recommended AI capabilities to reduce their effort. We also revealed tensions in the design of such an agent. We then turned to deliberative theory to examine the efficacy of deliberation. We argued that deliberation encourages reasoned dialog and critical thinking among participants, thus bringing value despite surfacing individual opinions and not facts. We then compared deliberation to other approaches such as moderation and fact-checking, discussing how deliberation can help increase the efficacy of those existing infrastructures. We concluded by distilling design recommendations, limitations, and future work.

% First, we discussed the deliberative environment. Participants suggested three activation methods: heuristics, AI-based, and manual activation, with a preference for manual activation in smaller groups due to privacy concerns. The debate on participation revolved around reaching all members versus a random subset, while the consensus for the duration was waiting one day to accommodate diverse schedules. Then, we discussed the process. Participants favored researching before giving opinions but expressed concerns about workload. They preferred anonymity in the summary message and were ambivalent about content moderation based on deliberation outcomes. Finally, we discussed the outcomes, focusing on the strengths and weaknesses of the process. Anonymous deliberation encouraged open discussions and diverse perspectives but could disrupt group dynamics. Some participants preferred simple content flagging and were unsure about its efficacy for extreme users

\begin{acks}
We sincerely thank the participants for sharing their deep insights into the role AI technologies can play in addressing harmful content. We also extend our gratitude to the Cornell Center for Social Sciences, the Mario Einaudi Center for International Studies, and Global Cornell for supporting this work.
\end{acks}

\bibliographystyle{ACM-Reference-Format}
\bibliography{references}

\appendix

\includepdf[pages=1, scale=.8, pagecommand={\section{Design Worksheet}\label{sec:design_worksheet}\thispagestyle{plain}}]{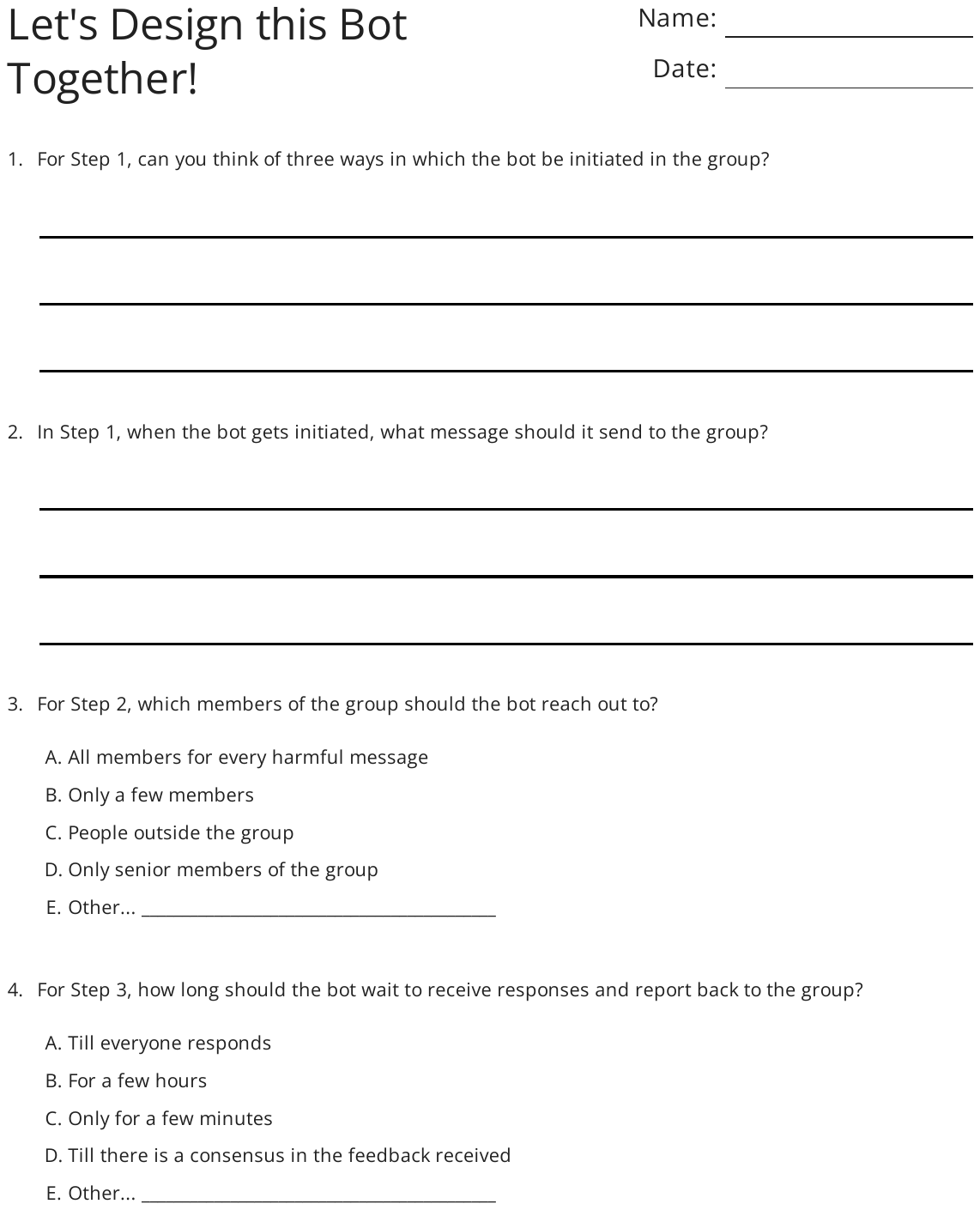}
\includepdf[pages=2-, scale=.8, pagecommand=\thispagestyle{plain}]{images/Design_Worksheet.pdf}

% \section{Conversational Agent Screenshots}
% Please refer Figure~\ref{fig:bot_design}.

% \begin{figure}[t]
%     \centering
%     \begin{tabular}{ccc}
%     \includegraphics[height=2.4in]{images/slide3.png} & \includegraphics[height=2.4in]{images/slide4.png} & \includegraphics[height=2.4in]{images/slide5.png} \\
%     (a) Activation & (b) Polling opinions & (c) Summarization  \\
%     \end{tabular}
%     \caption{Screenshots of the three phases in which the agent functions.}
%     \label{fig:bot_design}
% \end{figure}

\end{document}